\pdfoutput=1

\documentclass[twocolumn]{aastex62}

\usepackage{graphicx, latexsym, textcomp, gensymb, multirow, booktabs, amsfonts, amsmath, amssymb, natbib, url, hyperref}
\usepackage{threeparttablex}
\usepackage{ulem}

\usepackage[space]{grffile}

\newif\iflatexml\latexmlfalse

\usepackage[utf8]{inputenc}
\usepackage[english]{babel}

\newcommand{\kms}{km s$^{-1}$}
\newcommand{\kpc}{\ensuremath{\textrm{kpc}}}
\newcommand{\msun}{\ensuremath{\textrm{M}_\odot}}

\def\Msun  {${\rm M}_\odot$}

\def\arcmin {$^\prime$}

\def\kms   {km s$^{-1}$}
\def\Mhi   {M$_{\rm HI}$}

\def\ccm {cm$^{-3}$}
\def\cm {cm$^{-2}$}

\shorttitle{Local Group Dwarf HI}
\shortauthors{Putman et al.}

\begin{document}

\defcitealias{grcevich09}{GP09}
\defcitealias{spekkens14}{S14}
\defcitealias{simon19}{S19}
\defcitealias{mcconnachie12}{M12}

\title{The Gas Content and Stripping of Local Group Dwarf Galaxies}

\shorttitle{Gas Content in Local Group Dwarf Galaxies}
\shortauthors{Putman et al.}

\newcommand{\columbia}{Department of Astronomy, Columbia University, New York, NY 10027, USA}
\newcommand{\stsci}{Space Telescope Science Institute, 3700 San Martin Drive,  Baltimore, MD, 21218}
\newcommand{\jhu}{Department of Physics \& Astronomy, Johns Hopkins University, 3400 N. Charles Street, Baltimore, MD, 21218}
\newcommand{\cca}{Center for Computational Astrophysics, Flatiron Institute, Simons Foundation, 162 Fifth Avenue, New York, NY 10010, USA}
\newcommand{\berkeley}{Department of Astronomy, University of California, Berkeley, Berkeley, CA 94720, USA}
\newcommand{\miller}{Miller Institute for Basic Research in Science, University of California, Berkeley, Berkeley, CA 94720, USA}

\author[0000-0002-1129-1873]{Mary~E.~Putman}
\affiliation{\columbia}
\email{mputman@astro.columbia.edu}
\correspondingauthor{Mary~E.~Putman}

\author[0000-0003-4158-5116]{Yong~Zheng}
\affiliation{\berkeley}
\affiliation{\miller}

\author[0000-0003-0872-7098]{Adrian~M.~Price-Whelan}
\affiliation{\cca}

\author[0000-0002-6521-1920]{Jana~Grcevich}
\affiliation{\columbia}

\author{Amalya~C.~Johnson}
\affiliation{\columbia}

\author[0000-0002-9599-310X]{Erik~Tollerud}
\affiliation{\stsci}

\author[0000-0003-4797-7030]{Joshua~E.~G.~Peek}
\affiliation{\stsci}

\begin{abstract}
The gas content of the complete compilation of Local Group dwarf galaxies (119 within 2 Mpc) is presented using HI survey data.   Within the virial radius of the Milky Way (224 kpc here), 53 of 55 dwarf galaxies are devoid of gas to limits of \Mhi$~<10^4$~\Msun.  Within the virial radius of M31 (266 kpc), 27 of 30 dwarf galaxies are devoid of gas (with limits typically $<10^5$~\Msun).  Beyond the virial radii of the Milky Way and M31, the majority of the dwarf galaxies have detected HI gas and have HI masses higher than the limits. When the relationship between gas content and distance is investigated using a Local Group virial radius, more of the non-detected dwarf galaxies are within this radius (85$\pm1$ of the 93 non-detected dwarf galaxies) than within the virial radii of the Milky Way and M31. Using the \emph{Gaia} proper motion measurements available for 38 dwarf galaxies, the minimum gas density required to completely strip them of gas is calculated.  Halo densities between $10^{-5}$ and $5 \times 10^{ -4}$~\ccm~are typically required for instantaneous stripping at perigalacticon.  When compared to halo density with radius expectations from simulations and observations, 80\% of the dwarf galaxies with proper motions are consistent with being stripped by ram pressure at Milky Way pericenter.  The results suggest a diffuse gaseous galactic halo medium is important in quenching dwarf galaxies, and that a Local Group medium also potentially plays a role.

\end{abstract}


\section{Introduction}
\label{intro}

The smallest galaxies in the Universe today provide a window into how the first galaxies that formed in the Universe have evolved.  These small galaxies can only be detected locally and have been found to have a range of properties; from objects that appear to harbor only dark matter and just thousands of stars, to gas-rich dwarf galaxies with a range of stellar populations.  It remains unclear what dictates the properties of the dwarf galaxies, and in particular what quenches their gas content and star formation.  This understanding is key to make progress on how these small galaxies evolve and proceed to build up larger galaxies. 

In the Local Group, the gas content of dwarf galaxies has been shown to have a clear correlation with distance from the primary spiral galaxy (e.g., \citealt{grcevich09} (hereafter GP09), \citealt{spekkens14, einasto74, blitz00, grebel03}).
This suggests the environment of the dwarf galaxy plays an essential role in its evolution. Beyond the Local Group, however, the situation is not as clear. The SAGA survey of Milky Way analogs finds more star forming dwarf galaxies in the halos of most of these galaxies than found for the Milky Way or M31 \citep{geha17}. Other work finds more consistency with the Local Group, with evidence for the gas content of dwarf galaxies increasing with isolation \citep{stierwalt15, bradford15}, and a similar distribution of dwarf galaxies around spirals in a deep optical survey \citep{carlsten20}.

There are numerous mechanisms that can rob a dwarf galaxy of its cold gas. In the early universe, reionization is thought to be effective at removing cold gas in the smallest dark matter halos \citep{tollerud18, rodriguez19, kang19}. It is possible this mechanism is patchy in its effectiveness, and the exact redshift is still debated, but the star formation histories of some of the oldest local dwarf galaxies are consistent with this being a quenching mechanism \citep{weisz14}. Another mechanism is through some type of interaction with another galaxy. There is no strong evidence that a dwarf-dwarf interaction removes the cold gas from the resulting pair \citep{pearson16,pearson18}, but proximity of the pair to a larger galaxy is often found to result in quenched galaxies \citep{stierwalt15,geha12}. Supernova explosions and stellar winds can also act to remove gas, particularly during periods of intense star formation, but this appears to be insufficient for the complete removal of gas in typical systems \citep{emerick16, agertz13}.

In the Local Group, dwarf galaxies within the virial radius of the Milky Way or Andromeda are largely devoid of gas, while those beyond typically have gas.
This relationship suggests that the distance to a larger galaxy is an important factor in dwarf galaxy quenching.
Though tidal forces can be important for those galaxies that have plunged deep into the host dark matter halo in the past, this has been shown to affect only a very small fraction of the Local Group dwarfs \citep[e.g.,][]{mateo08,simpson18,iorio19}. 
The results thus far have been consistent with a diffuse gaseous halo medium playing an important role in stripping the gas from the dwarf galaxies given the range of distances for the dwarf galaxies and the assumption of instantaneous stripping \citep[GP09,][]{gatto13}. This can be tested with the discovery of new dwarf galaxies, further information about the orbital histories of the dwarf galaxies, and deeper limits on the HI content of the dwarf galaxies.  
 
One aspect that has not been commonly considered is the role of a diffuse group medium of the Local Group in stripping dwarf galaxies. Groups are thought to harbor a hot halo medium at temperatures of $\sim10^{5-6}$~K \citep{osmond04,nuza14,stocke19},  
and dwarf galaxies falling through this medium with a fast enough velocity are likely to be partially stripped. Simulations also show support for some dwarf galaxies being more easily quenched in a Local Group environment compared to the equivalent satellite of an isolated spiral \citep{garrison19}. 
However, the role of large scale environment, instead of that of an individual host galaxy, in dictating a galaxy's gas content continues to be debated \citep{luber19, tollerud16}. 
 
In this paper, we provide new limits on the gas content of the Local Group dwarf galaxy population for  galaxies discovered within 2 Mpc as of 2020 (\citealt{simon19}, hereafter S19; \citealt{mcconnachie12}, hereafter M12). This work is motivated by the discovery of large numbers of new Local Group dwarf galaxies, the availability of deeper HI survey data, and the uncertainty on the cause of dwarf galaxy quenching.  Gas content can be used to describe the HI results as molecular and ionized gas are only detected in those galaxies with HI and are generally a small fraction of the HI \citep{schruba12,buyle06,gallagher03}.  In \S2 we describe the data and methods we use to set the HI mass limits.  A model for the structure of the Local Group is presented in \S3.1, and the Milky Way model used for orbit integration of the dwarf galaxies is described in \S3.2. In \S4 we present the HI results in the context of the distance to the Milky Way or Andromeda and relative to a virial radius of the Local Group.  We discuss our results in \S5 in terms of the recent \textit{Gaia} proper motion measurements for Milky Way dwarf galaxies and the potential role of a Local Group medium. The paper is briefly summarized in \S6.

\section{Data and HI Mass Limit Derivations} 
\label{sect:data}

The dwarf galaxies included in this study are listed in Table~\ref{tab:hi} sorted by their heliocentric distance. This distance is also used to calculate the HI mass limit. Additional properties of the dwarf galaxies are listed in Table~\ref{tab:other}. The dwarf galaxies and their properties are primarily taken from the updated online catalog\footnote{\url{http://www.astro.uvic.ca/~alan/Nearby_Dwarf_Database.html}} of the M12 paper and the S19 review.  The online M12 catalog contains those dwarf galaxies within 3 Mpc that have distances measured from resolved stellar populations.  Here we include only those galaxies within 2 Mpc to aim for greater completeness and to avoid nearby groups. This catalog includes some galaxies that are clearly not dwarf 
 galaxies, so we exclude
 galaxies with stellar masses $>2\times10^9$~\msun.  Note this limit results in the \emph{inclusion} of the Magellanic Clouds.  We also exclude Canis Major from the M12 list, as it is primarily defined as a stellar stream \citep{martin04}.  We add to the list additional ultra-faint dwarf galaxies tabulated in S19 and Antlia 2 \citep{torrealba19}.  For all of the ultra-faint dwarf galaxies, we prioritize the more recent S19 values for their properties over M12.  We examined the catalog of \cite{kara14} as a check of our dwarf galaxy completeness within 2 Mpc and found anything additional in their catalog to be consistent with being unconfirmed or a globular cluster.  

There are 3 main sources of HI data we used to set limits on the dwarf galaxies HI content: GALFA-HI DR1 \citep{peek11}, GALFA-HI DR2\footnote{\url{https://purcell.ssl.berkeley.edu/}} \citep{peek18}, and HI4PI \citep{hi4pi}.  These are referred to as GALFA1, GALFA2 or HI4PI in Table 1.  The starting point for all of the GALFA-HI (Galactic Arecibo L-band Feed Array HI) data were cubes with a resolution of 4\arcmin~and a channel spacing of 0.754~\kms.  GALFA-HI DR2 covers the entire sky observable with the Arecibo 300-m radio telescope, or all right ascensions for declinations between approximately $-1$ to $+37$ degrees (approximately 1/3 of the sky). GALFA-HI DR1 has the same resolution as DR2, but covers only approximately half of the Arecibo sky in non-continuous regions.  We used the DR1 data in a few cases because they are deeper in some regions with the inclusion of additional commensal data \citep[see][]{peek11,peek18}. 
The HI4PI survey combines the southern GASS data taken with the Parkes 64-m Radio Telescope \citep{mccluregriffiths09} and the northern EBHIS survey taken with the Effelsberg 100-m telescope to cover the entire sky \citep{winkel16}.  The combination results in an all-sky HI survey with 16.2\arcmin~spatial resolution and a channel spacing of 1.29~\kms.

To determine if a dwarf galaxy was detected and to obtain the limits on the HI masses, we first applied Hanning smoothing on each cube that contains the position of a dwarf galaxy to $\sim10$~\kms.  Since the channel spacing is different for GALFA-HI and HI4PI, the closest value to 10~\kms~was used, 13 channels were smoothed for GALFA-HI and 7 channels for HI4PI.   We then examined the smoothed data cube at the position and velocity of the source in both the data cube and the spectrum for a possible detection.  In all cases where there was no previous detection, there was no detection at the position and velocity of the galaxy and we proceeded to extract a noise level to determine the HI mass limit.

The method to obtain the quantitative HI mass limit differed depending on whether the gas in the dwarf was likely to be resolved or not.  We used the optical half-light radius ($r_h$) as a guide to whether this was likely.  In all cases the noise values were checked to be consistent across several channels and, with the exception of the rare cases when the dwarf is resolved with HI4PI data, the standard deviation was calculated within a 30\arcmin~diameter region at the position and velocity of the dwarf.  The 30\arcmin~window is designed to be an area larger than the beam size, but not large enough to potentially encompass nearby structures at the velocity of the dwarf (i.e., Galactic emission in some cases or high velocity clouds).  None of the dwarf galaxies were close enough to the edges of the cubes for this to influence the analysis.  If the dwarf is in the GALFA-HI sky and the optical diameter is smaller than 4--5\arcmin, the limit was obtained from the standard deviation within a 30\arcmin~diameter region at the position and velocity of the dwarf (Unres.~in Table~\ref{tab:hi} with Data=GALFA1 or 2). If the dwarf is in the GALFA-HI sky and has an optical diameter between approximately 5--12\arcmin~we do this same procedure with GALFA-HI data smoothed to 8\arcmin~(Res.~in Table~\ref{tab:hi} with Data=GALFA1 or 2).  This generally resulted in a better limit on the HI mass than going to the HI4PI data, though for the three dwarf galaxies with optical diameters $>9$\arcmin~it does assume the HI is more centrally concentrated (Coma Berenices, Andromeda XIX and Andromeda II).  There are several dwarf galaxies that we adopt the HI4PI limit even though the dwarf is in the GALFA-HI sky: Hercules and Canes Venatici I because of their large optical size, and Canes Venatici II, Andromeda XI, Leo I and Leo II due to their location in a noisier region of the GALFA-HI data (i.e., on a basketweave remnant and/or near the outer regions of the survey).  

HI4PI data was used for the majority of the dwarf galaxies due to its greater sky coverage.   
The limits were obtained with a 30\arcmin~region (Unres.~in Table~\ref{tab:hi} with Data=HI4PI), as was used for the GALFA data, with the exception of those galaxies that have optical sizes larger than the 16.2\arcmin~HI4PI beam.  For these galaxies we took the sum of the flux within a region the optical size of the galaxy at its position and did the same for 8 regions surrounding the central position of the galaxy.  We divided the summed flux values by the area of the optical size and calculated the standard deviation of the 9 values.  We then check the value is consistent with values taken from surrounding channels (Res.~in Table~\ref{tab:hi} with Data=HI4PI).   
The exceptions to using this method for large extent galaxies are the disrupted Sagittarius dSph and Bootes III galaxies.  For these galaxies we quote the limit in the core with the unresolved technique rather than attempt to cover their full extent.
The $1\sigma$ values from these methods are listed in Table 1, while $5\sigma$ is used in the HI mass calculations.  The conversion factors from the Kelvin $\sigma$ values to Jy are 0.6 Jy/K for HI4PI, 0.11 Jy/K for GALFA-HI, and 0.44 Jy/K for the GALFA-HI data smoothed to 8\arcmin~resolution.

Several dwarf galaxies are at velocities that are contaminated by Galactic emission.  In most cases, five times the standard deviation calculated with the relevant above method is sufficiently high to represent a limit on the HI gas in the dwarf over the background. In the case of Willman 1, its low heliocentric velocity led to strong Galactic emission and the standard deviation not being suitable for determining the HI mass limit.  For this galaxy we calculated the median flux in 30\arcmin~regions surrounding the dwarf position in the 10 channels centered on the dwarf velocity and found twice the median of these values was a conservative value to choose for a limit on what would be a detectable HI mass.

Ultimately, $5\sigma$ HI mass limits are quoted in Table~\ref{tab:hi} and they are calculated using a velocity width of 10~\kms~and the distance to the dwarf galaxy: M$_{\rm HI}~(\msun) = 2.36 \times 10^5$ $5\sigma$(Jy) 10~\kms~D$^2$(Mpc).   We debated whether to use the stellar dispersion values for each galaxy instead of a fixed 10~\kms, but given the variations on availability and errors we chose the fixed value.  Errors on the mass limits are given based on the distance errors.
If the systemic velocity of the dwarf galaxy was not available we measured the standard deviation every 10~\kms~for $|v_{\rm LSR}| < 500$~\kms~in a 30\arcmin~region at the position of the dwarf (all are unresolved and labeled Unres. No V) and used the average of those values, excluding those channels that were contaminated by Galactic or Magellanic-related emission.  
The dwarf galaxies without velocities at the time of paper submission were:  Cetus II, Horologium II, Reticulum III, Pictoris 1, Columba I, Sagittarius II, Indus II, Phoenix 2, Eridanus 3, Indus I, Cetus III, DESJ0225+0304, Pictor II, and Virgo I.

Though we checked some of the HI masses for previously detected dwarf galaxies for consistency, we did not remeasure the HI masses and largely adopted values from the references in M12 and adjusted the value if an updated distance was available.  
The only exceptions to adopting the M12 detections are Sculptor and Fornax.  We do not include them because the HI clouds detected are not centered on the optical component of the galaxies and there is abundant nearby HI that makes the association highly uncertain. Instead, we adopt HI limits at the stellar position of these galaxies.  Sculptor formed the vast majority of its stars in the distant past and is in a region where there are multiple HVCs; therefore we think it is unlikely the nearby gas is associated.   Fornax has formed stars in the past few Gyrs \citep{weisz11}, but its velocity makes confusion with Galactic HI emission a huge issue and the potentially associated extended HI is not centered on the galaxy.  
The Phoenix dwarf galaxy was considered highly uncertain as well, but with the improved stellar velocities of \cite{kacharov17} and the HI clump at the stellar heliocentric velocity and overlapping with the stellar component \citep{stgermain99, young07}, we adopt this HI clump as a detection.

\begin{figure*}[t]
\centering
\includegraphics[width=1\textwidth]{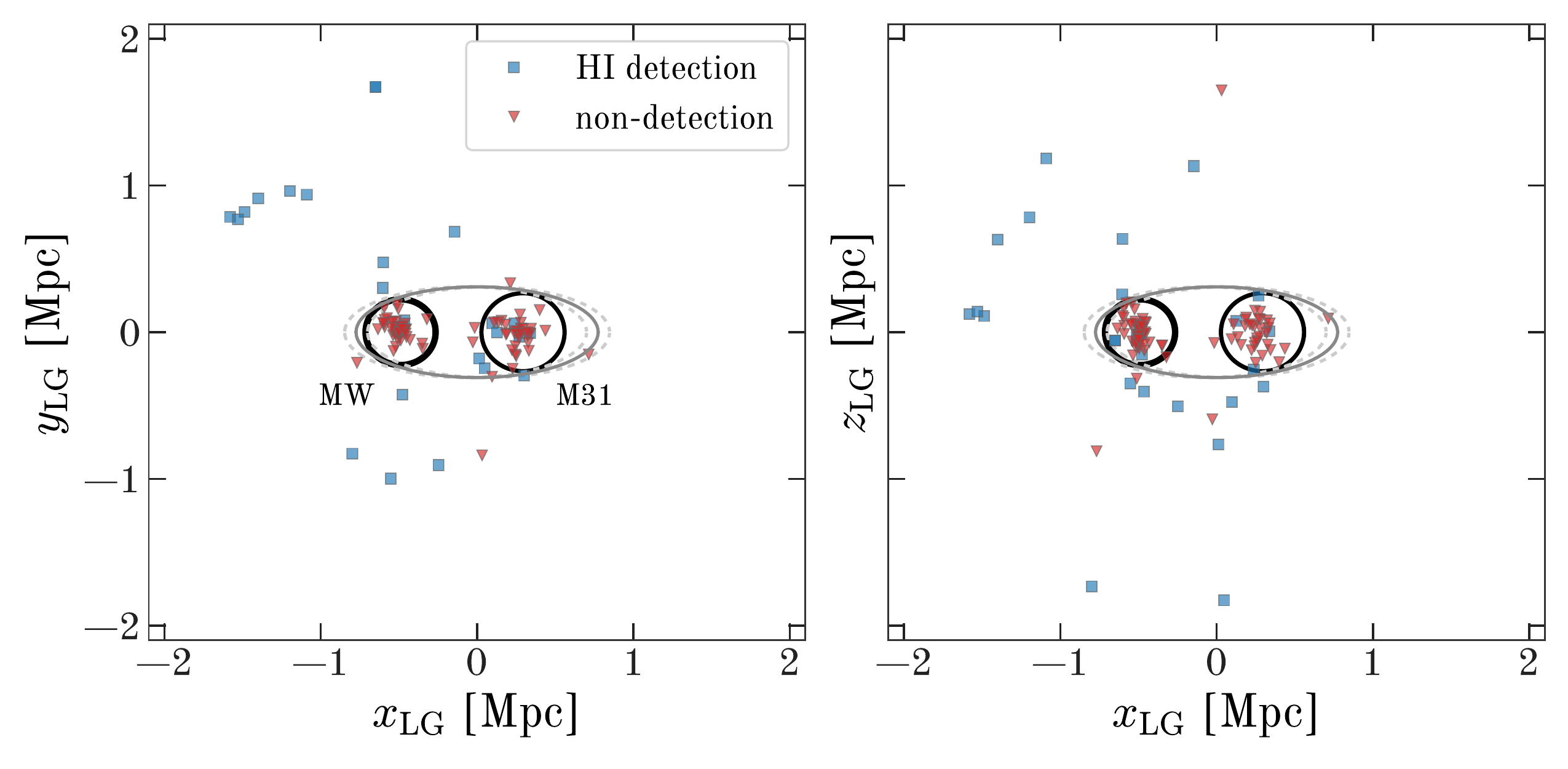}
\caption{Schematic of the Local Group boundary surface or approximate virial radius of the Local Group (solid, gray line) and positions of dwarf galaxies in our sample (markers), shown in two Cartesian projections (left and right panels) of the LG coordinates defined in Section~\ref{sect:lgsurf}.
The dashed gray ellipses show the LG boundary surfaces in two alternate LG mass models that adjust the masses of the MW and M31 (see Section~\ref{sect:lgsurf}).
The square (blue) markers show galaxies with detected HI and upside-down triangle (red) markers show non-detections.
In each panel, the large (black) annuli show the range of virial radii of the Milky Way (left) and M31 (right) models considered.}
\label{fig:lgsurf}
\end{figure*}

\section{Local Group and Milky Way Mass Models}
\label{sect:model}

In this section, we define mass models for the Milky Way and M31 that we use to define a Local Group virial radius or `surface'.  We then describe how we numerically computed orbits for the satellites of the Milky Way that have proper motion measurements.

\subsection{Defining a Local Group `Bounding Surface'}
\label{sect:lgsurf}

We first define a Local Group (LG) coordinate system with an origin at the approximate center-of-mass of the Milky Way and M31 (computed using models specified below).
We define the $x$-axis of this coordinate system by the line connecting the Galactic Center to the center of M31 and place the origin at the center-of-mass along this axis.
Because our models are symmetric about the $x$ axis, the orientation of the $y$ and $z$ axes are arbitrary; with our definition, the $z$-axis is aligned with $(\alpha, \delta)_{\rm ICRS} = (193.327, 48.585)^\circ$, and the $y$-axis with $(\alpha, \delta)_{\rm ICRS} = (102.219, 0.977)^\circ$. 
We assume the heliocentric distance to M31 is $779~\kpc$ \citep{Conn:2012}, the heliocentric distance to the Galactic Center is $8.122~\kpc$ \citep{GRAVITY}, and the coordinates of the Galactic Center are taken from \cite{Reid:2004}.

As a simplified representation of the total density distribution of the LG, we define a prolate spheroidal surface bounding the LG in order to distinguish satellite systems that are within or outside of the LG.
We first construct a mass model for the total mass distribution of the Milky Way and M31 by representing each galaxy with a spherical Navarro-Frenk-White (NFW; \citealt{nfw}) density profile such that the total LG density is given by the sum of two independent NFW profiles.
For the Milky Way, we use virial mass and halo concentration values of $M_{200, \textrm{MW}}=1.2 \times 10^{12}~\msun$ and $c=10$, motivated by a number of recent mass determinations that use satellite galaxy and globular cluster kinematics to infer the mass of the Galaxy \citep[e.g.,][]{Eadie:2019, Posti:2019, Wegg:2019}. 
For M31, we adopt $M_{200, \textrm{MW}}=2 \times 10^{12}~\msun$ \citep{Fardal:2013} and use the same concentration parameter as the Milky Way.
The corresponding NFW scale radii are therefore $r_{s, \textrm{MW}}=22~\kpc$ and $r_{s, \textrm{M31}}\approx27~\kpc$, respectively.
We note that these values are also motivated by and consistent with the measurement of the total Local Group mass $M_{\rm LG} \approx 3.2\times10^{12}~\msun$ \citep{VanDerMarel:2012}.

To determine the axis ratio of the total LG density distribution, $q_{\textrm{LG}}$, we compute the $xx$ and $yy$ components of the inertia tensor given by the above mass model and use these to compute the axis ratio $q_{\textrm{LG}} = I_{yy} / I_{xx}$.
We define the LG-bounding prolate surface (in LG Cartesian coordinates $x,y,z$) to be 
\begin{equation}
    x^2 + \frac{y^2 + z^2}{q_{\textrm{LG}}^2} = r_e^2
\end{equation}
where $r_e \approx 733~\kpc$ is defined such that the surface fully encloses the virial sphere around M31. 
For each satellite galaxy in our sample, we compute the minimum distance between the galaxy and the LG-bounding surface.
Figure~\ref{fig:lgsurf} (solid gray ellipse) shows a projection of this bounding surface in both the $x_{\textrm{LG}}, y_{\textrm{LG}}$ and $x_{\textrm{LG}}, z_{\textrm{LG}}$ planes, along with the positions of all dwarf galaxies in our sample (blue squares for HI detected-galaxies and red upside-down triangles for non-detections) and projections of the virial spheres around the Milky Way and M31 (left and right black circles, respectively).

We note that there is substantial uncertainty and potentially large (up to a factor of $\sim$2) biases on the measurements of both the Milky Way and M31 virial masses \citep[e.g., from the presence of the Magellanic Clouds in the case of the Milky Way,][]{Erkal:2020}, so this bounding surface should be interpreted qualitatively as a ``soft'' boundary.
To emphasize this point, and to assess how our conclusions and results are affected by uncertainties on the global properties of the MW and M31 dark matter halos, we also construct two additional LG models and bounding surfaces with varied masses for the two galaxies.
Since the total LG mass is fairly well constrained \citep{VanDerMarel:2012}, we only vary the masses of the MW and M31 to keep the total LG mass constant at $M_{\rm LG} = 3.2\times10^{12}~\msun$.
As the virial masses of the MW or M31 are not individually well constrained, and the scatter in the existing mass measurements for either galaxy are likely under-estimated due to biases in the individual measurements \citep[e.g.,][]{Kafle:2018}, we choose to vary the MW mass by 25\% up and down (adjusting the mass of M31 to keep the LG mass fixed).
Our models therefore have $M_{\rm MW} = (0.9, 1.2, 1.5) \times 10^{12}~\msun$ and $M_{\rm M31} = (2.3, 2.0, 1.7) \times 10^{12}~\msun$, $R_{{\rm vir}, MW} \approx (204, 224, 242)~\kpc$, and $R_{{\rm vir}, M31} = (279, 266, 252)~\kpc$, respectively.
The LG bounding surfaces for the two other models are shown in Figure~\ref{fig:lgsurf} as dashed ellipses (gray), and the thickness of the virial circles reflects the range of virial radii for the two galaxies.

In Table~\ref{tab:other}, we include estimates of the closest distance between each dwarf galaxy and the LG surface, $D_{\rm LG}$, as estimated in our fiducial mass model.
We use the two other mass models to estimate bounds on the $D_{\rm LG}$ values for each source and note these as error bars. 
We note that in some cases (e.g., Pisces~II, Pegasus~III), the LG surface distance in the fiducial mass model is not the central value; this is expected, and happens when a dwarf galaxy is near enough to the LG bounding surface such that the change of ellipticity of the surface in the different LG mass models causes unintuitive changes.

\subsection{A Milky Way model for orbit integration}
\label{sect:mwmodel}

We numerically integrate orbits for the Milky Way dwarf galaxies that have distances, radial velocities, and proper motion measurements.
For computing these orbits, we use a three-component mass model to represent the Milky Way composed of a Hernquist bulge \citep{Hernquist:1990}, Miyamoto-Nagai disk \citep{Miyamoto:1975}, and spherical NFW halo \citep{Navarro:1997}.
We fix parameters of the disk and bulge models to be consistent with the model defined in \citet{Bovy:2015}. 
For our fiducial model, we adopt a more massive halo such that the enclosed mass within $250~\kpc$ is $M_{\rm enc} \approx 1.2\times 10^{12}\,\msun$ in order to fit a compilation of mass enclosed measurements (see the \texttt{gala} documentation\footnote{\url{http://gala.adrian.pw/en/latest/potential/define-milky-way-model.html} } for more information) and to be consistent with the model assumed to define the LG boundary surface above.

The mass model is implemented in \texttt{gala} \citep{gala}, which we then also use to numerically integrate the orbits of the dwarfs, treating the satellite galaxies as test particles.
We cross-match our catalog of Milky Way satellites to recent proper motion measurements of a subset of the dwarfs \citep{fritz18, pace19} that make use of astrometry from \textit{Gaia} Data Release 2 \citep{Gaia-Collaboration:2018, Lindegren:2018}. 
We compute initial conditions (i.e., at present day) by generating random samples from the error distributions over all kinematic measurements for each of the 38 dwarf galaxies in this subset. 
For each error sample, we additionally vary the MW halo enclosed mass by sampling a uniform random value  between $0.9$--$1.5\times 10^{12}~\msun$, following the discussion in Section~\ref{sect:lgsurf}, in order to include uncertainty in the MW mass into our derived orbital quantities.\footnote{We note that this is likely overly conservative (sampling from a uniform distribution), but in any case we find that the uncertainties on the orbital quantities are dominated by measurement uncertainties on the galaxy kinematics.}
We transform these samples into a Galactocentric reference frame assuming a Sun-Galactic center distance of $R_\odot \approx 8.122~\kpc$ \citep{GRAVITY} and using solar motion values from \cite{Drimmel:2018}.
We then numerically integrate each orbit sample for each dwarf backwards from present day for $2~{\rm Gyr}$ with a timestep of $1~{\rm Myr}$ using Leapfrog integration.
We ignore the gravitational influence of the Magellanic Clouds, which likely affects the orbits of a small subset of the dwarf satellites of the Milky Way \citep{erkal20, patel20}.
We use these orbits (in particular, pericentric distances and velocities at pericenter) later to determine the minimum Milky Way gas density needed to strip the satellites of their own gas reservoirs.  Table~\ref{tab:orbit} includes orbital parameters for the satellites for which this is done.

\begin{figure*}
\centering
\includegraphics[width=\textwidth]{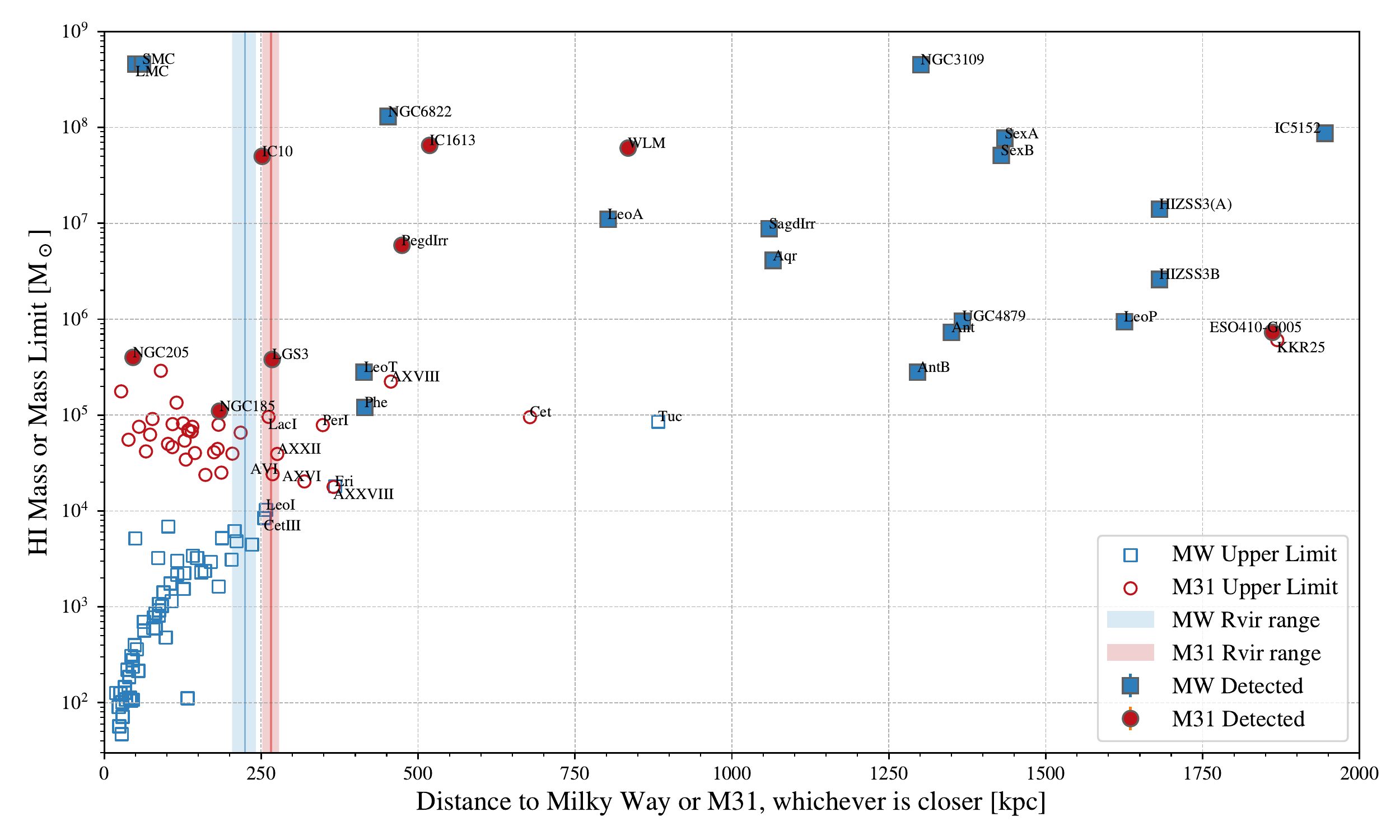}
\includegraphics[width=\textwidth]{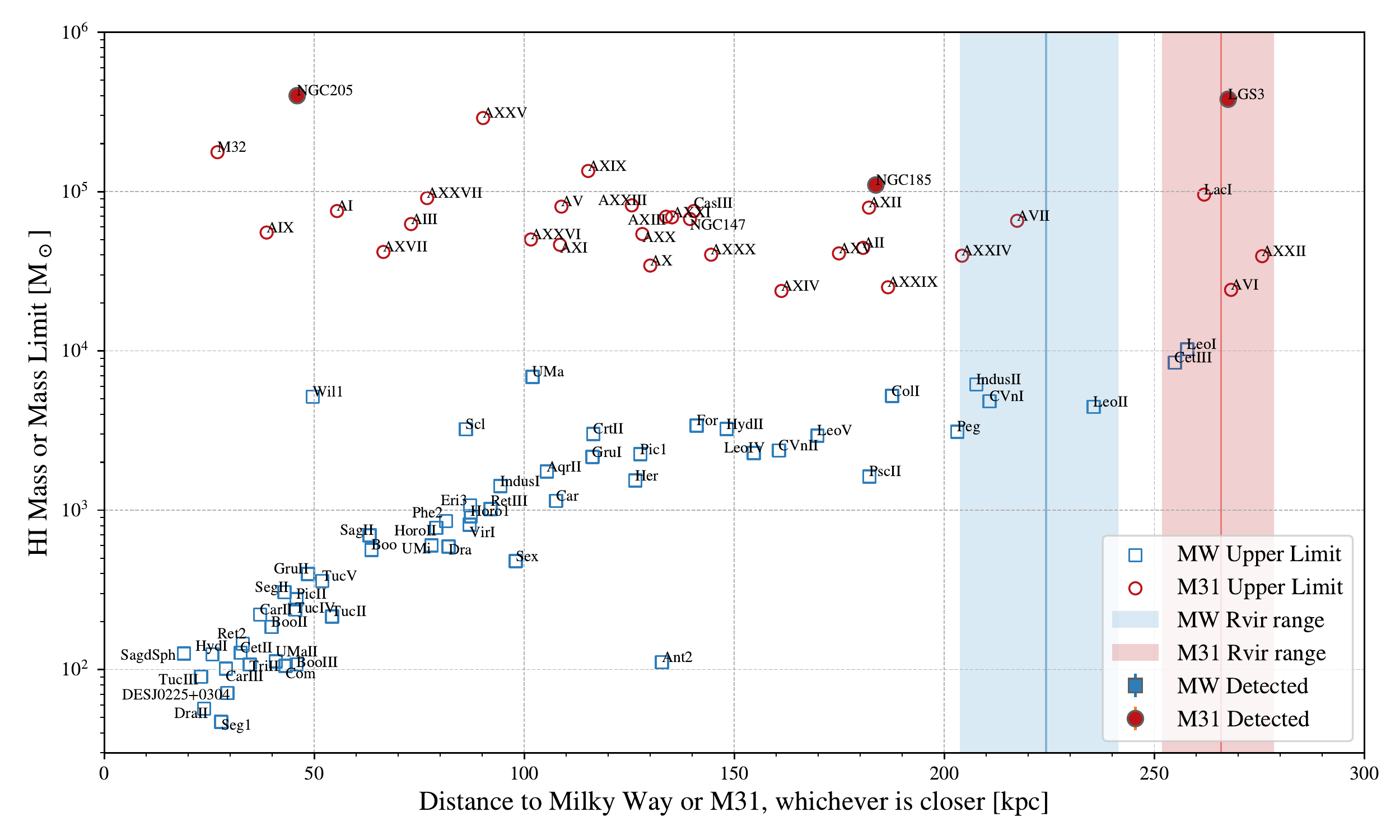}
\caption{\small The HI content of the dwarf galaxies (log scale) as a function of distance from the Milky Way (blue boxes) or M31 (red circles), whichever is closer.  $5\sigma$ limits on the HI content at the central position of dwarf galaxy are open symbols (labeled beyond 250 kpc of each galaxy in top plot), while detections are solid symbols (all labeled). The errors on these mass limits (from the distance errors) are smaller than the symbols.  
The blue line is at 224 kpc and represents the virial radius of the Milky Way with the shaded region representing the virial radii for 25\% larger and smaller masses.  The red line represents the virial radius of M31 at 266 kpc and the shaded region shows the virial radius range with the variation in MW mass and our fixed mass Local Group model. The bottom plot zooms in on the region within 300 kpc and for HI masses and limits $<10^6$~\msun, and all galaxies are labeled. }
\label{himwm31}
\end{figure*}

\section{Results}
\label{sect:results}

\begin{figure*}
\includegraphics[width=\textwidth]{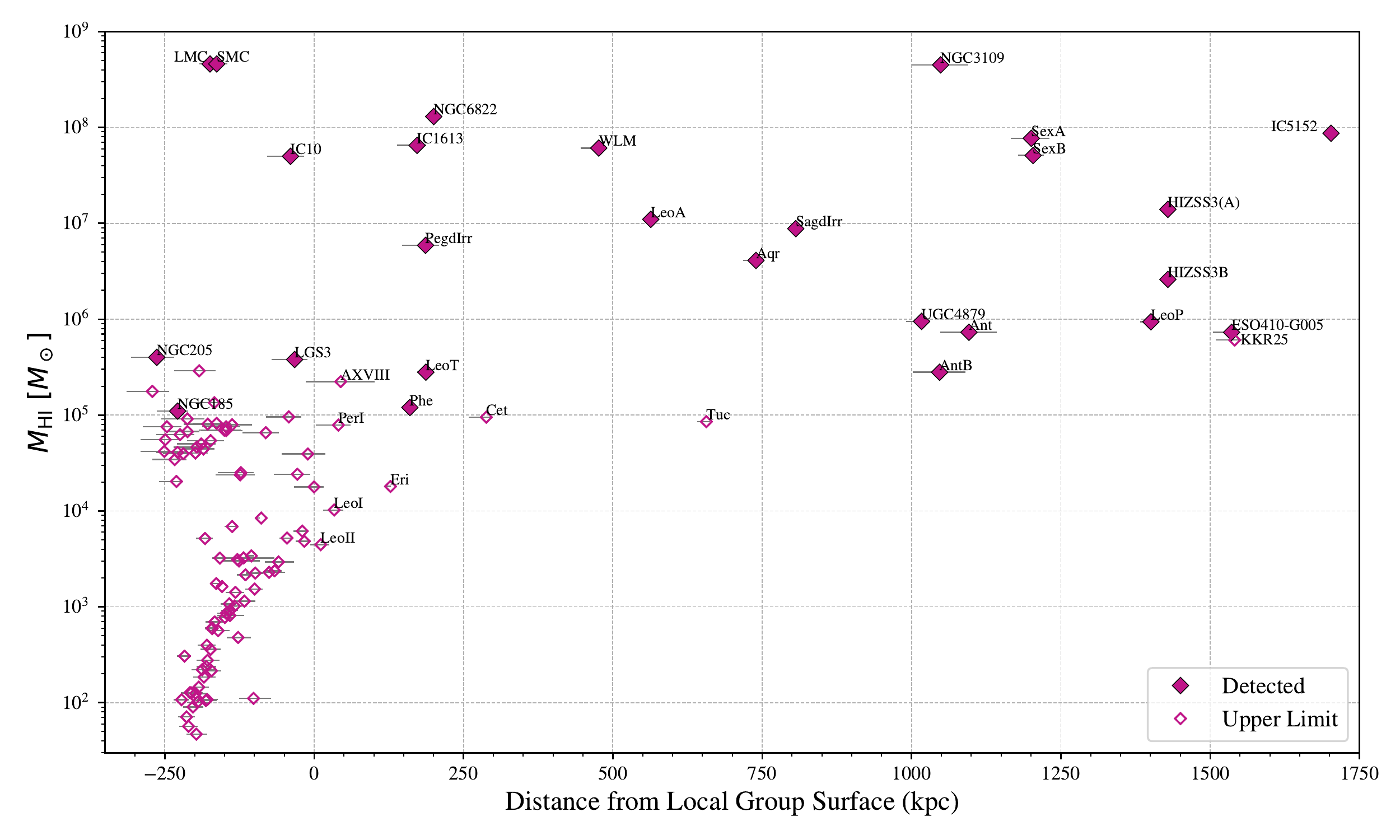}
\includegraphics[width=\textwidth]{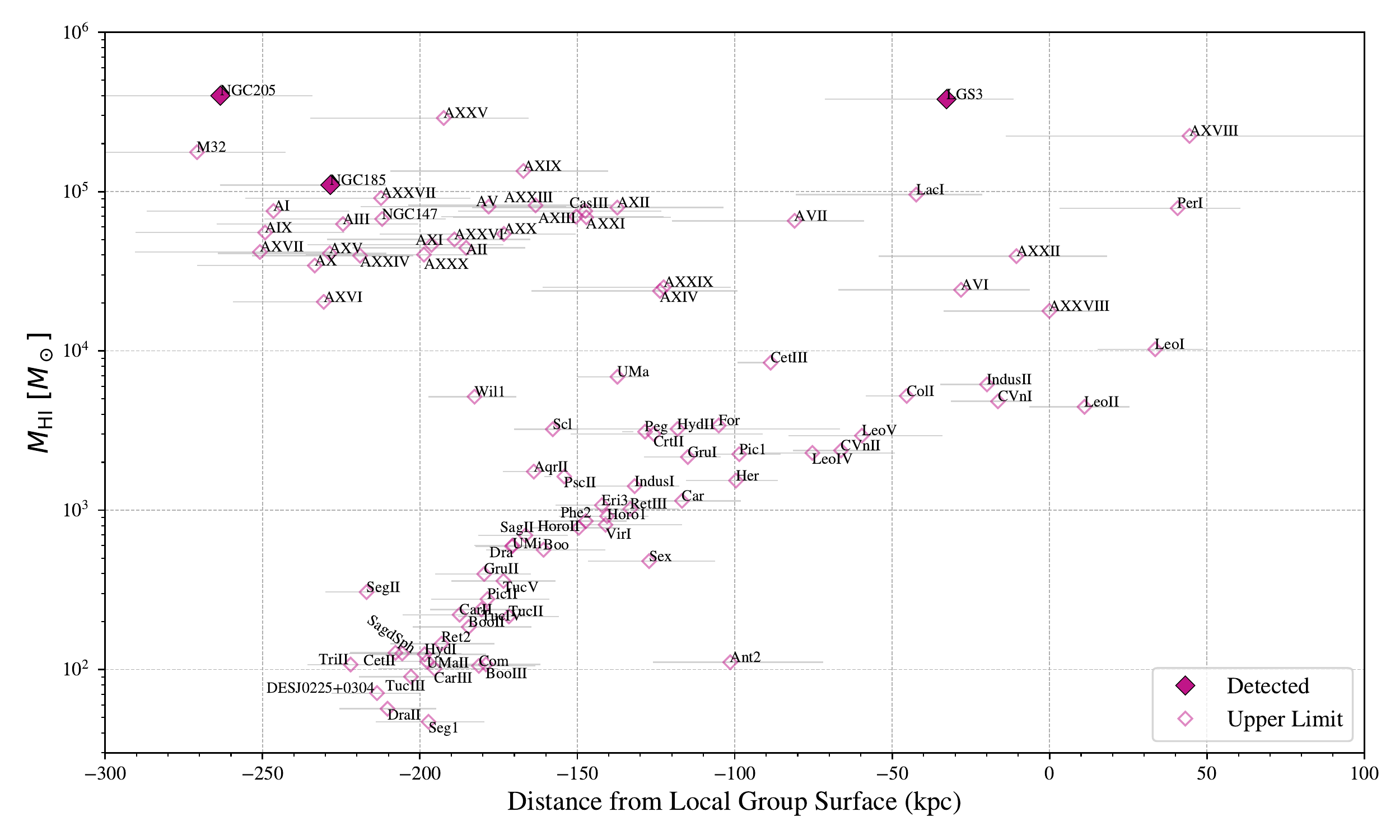}
\caption{\small The HI content of the dwarf galaxies (log scale) as a function of distance from a Local Group surface (at 0), as defined in \S\ref{sect:lgsurf}. The solid diamonds are HI detections and the open diamonds are $5\sigma$ HI mass limits.  All galaxies beyond the Local Group surface are labeled and only detections within the surface are labeled in the top plot.  The bottom plot zooms in on the inner region and all galaxies are labeled.   The error bars on the HI masses are smaller than the symbol sizes.  The horizontal bars on the Local Group surface distances are the range of possible distances from the Local Group surface using the variations in Milky Way and M31 mass described in \S\ref{sect:lgsurf}.}
\label{hilg}
\end{figure*}

The results of the new HI measurements are presented in Table~\ref{tab:hi} and shown in Figures~\ref{himwm31}--\ref{hilg}.  All of the newly discovered dwarf galaxies without previous HI measurements are non-detections, and all previous non-detections remain as non-detections.  The limits presented here are generally deeper than those previously published.  Notable other limits are the deeper limits published for some of the Milky Way dwarf galaxies in \cite{spekkens14} and pre-optical velocity limits for some of the new ultra-faint dwarf galaxies by \cite{westmeier15}. To ensure consistency across the complete set of Local Group dwarfs, our limits here depend only on the GALFA-HI and HI4PI data described in \S\ref{sect:data}.

Figure~\ref{himwm31} shows the relationship between the HI mass limit or HI mass of a dwarf galaxy versus distance from the Milky Way or M31, whichever is closer.  The virial radii of the Milky Way and M31, as defined in \S\ref{sect:model}, are 224 and 266 kpc, respectively. These virial radii are not presumed to be exact, but give a guideline to consider the detections and non-detections. A range of possible virial radii for each galaxy are shown by the shaded regions as described in \S~\ref{sect:lgsurf}.  The MW and M31 are described individually below, but it can be seen from these plots that the vast majority of the galaxies are non-detections within these radii, and beyond the vast majority are HI detections.  Though we code each dwarf galaxy by whether the MW or M31 is closer, we note that beyond a certain radius a clear association with either galaxy is uncertain without orbit information.  Thus, the majority of the HI detections cannot be directly associated with either the Milky Way or M31.  Beyond the virial radii of these galaxies, the detections Leo T, Phoenix and NGC6822 could be associated with the Milky Way, and LGS3, Pegasus dIrr and IC1613 with M31.  The other 15 detections beyond the virial radii in this plot are at $> 750$ kpc from either galaxy. The average HI mass of the 26 detected dwarf galaxies within 2 Mpc of the Milky Way is $7.5 \times 10^7$ \msun, and the median HI mass is $7.4 \times 10^6$ \msun.

Using the virial radius of 224 kpc for the Milky Way, 53 dwarf galaxies have $5\sigma$ limits on their HI mass of $<10^4$ \Msun~within this radius.  These are the strongest HI limits due to their proximity and the D$^2$ dependency for the HI mass.  This dependency also results in the arc of increasing mass limit with increasing distance since we use survey data that does not vary substantially in sensitivity at a given position.  For the MW dwarf non-detections beyond 224 kpc: Leo II has a HI mass limit $<10^4$ \Msun~and, at 233 kpc, is within the range of possible Milky Way virial radii we consider here, Leo I and Cetus III have HI limits $\sim10^4$ \Msun~and are at 254 and 251 kpc, and Eridanus II is at 366 kpc with an HI mass limit of $<2 \times 10^4$ \Msun. Tucana is a non-detection with a limit $<10^5$ \Msun~at 887 kpc from the Milky Way and should be considered a ``field'' object.  The lowest HI mass of a detected dwarf galaxy is above $10^5$ \Msun, so the results strongly indicate none of the MW dwarf galaxies with limits are likely to have HI.  The only 2 HI detections within the virial radius of the Milky Way are the Magellanic Clouds, which were not included in GP09 due to their large mass and are currently being stripped of their gas \citep{salem15,putman03}.  A radius of 260 kpc would optimize the numbers, with the MW having 56 dwarf galaxies with no gas, and 2 massive dwarf galaxies with gas within this radius.  We note that for Sculptor and Fornax, if the offset HI gas was associated (see \S\ref{sect:data}), they would be clear outliers in being small dwarf galaxies within 150 kpc of the MW that have HI gas.

\begin{figure*}
\includegraphics[scale=0.8,angle=0]{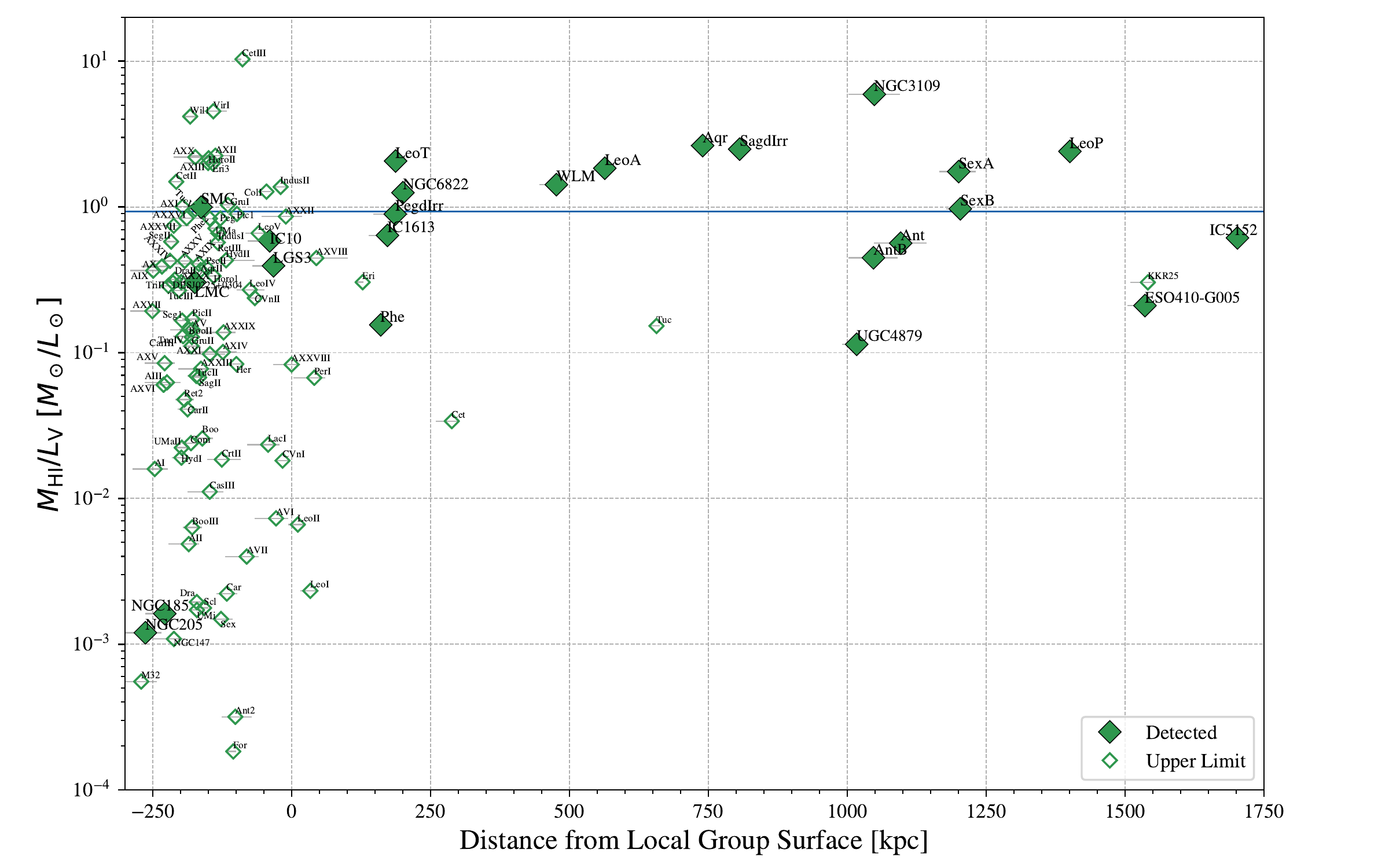}
\caption{\small The HI mass or mass limit of the dwarf galaxies divided by the V-band luminosity (log scale) against the distance from the Local Group surface.  Detections are solid diamonds and non-detections are open diamonds.  The lines through the symbols represent the range of possible Local Group surfaces by varying the mass of the MW and M31.  We exclude the Sag dSph from this plot as M$_{HI}/$L$_V$=$5.7 \times 10^{-6}$ due to its large V band luminosity and low HI mass limit.  The blue solid line represents the median M$_{HI}/$L$_V$ for the detected dwarf galaxies.}
\label{histar}
\end{figure*}

\begin{figure*}
\includegraphics[scale=0.8,angle=0]{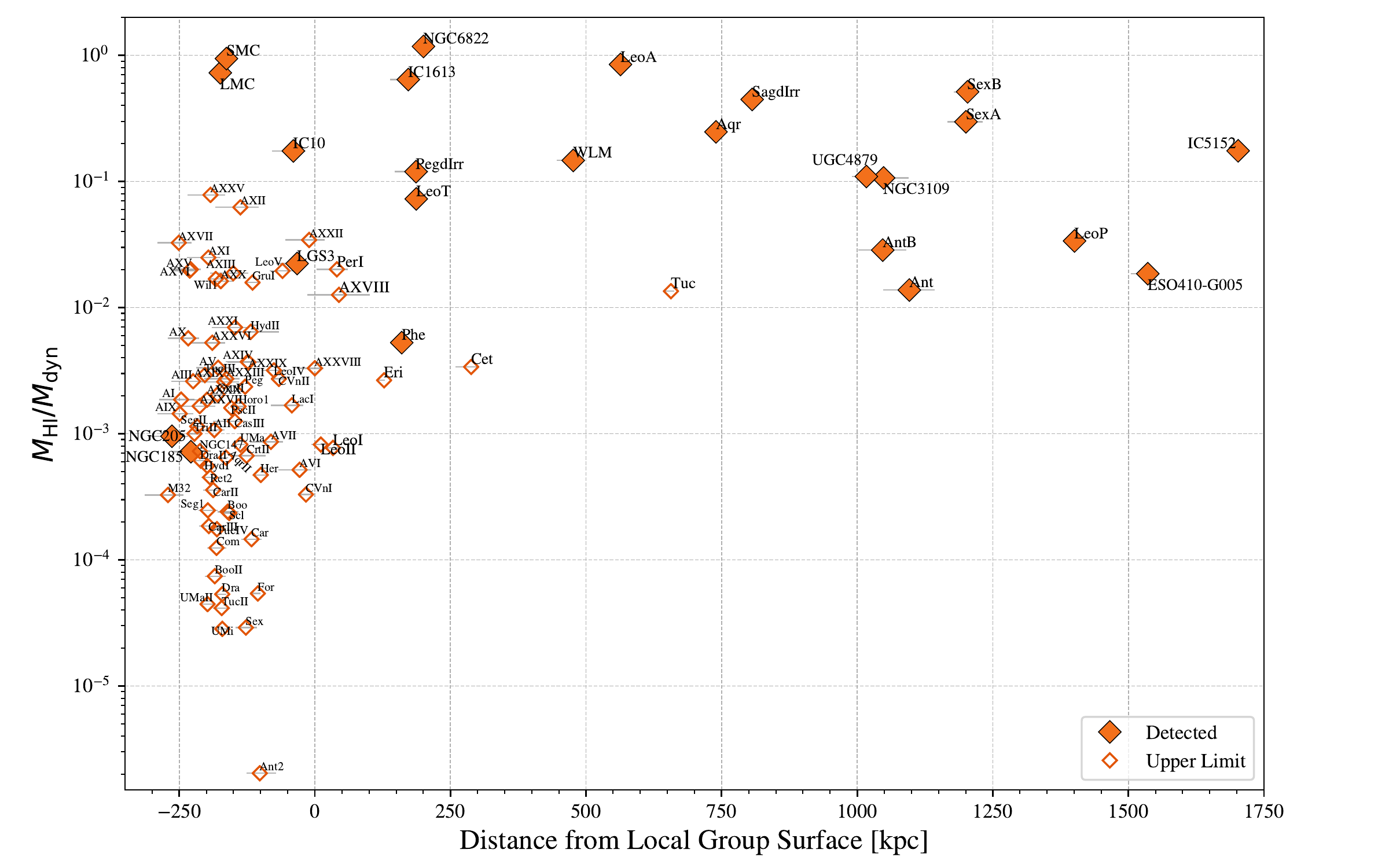}
\caption{\small The HI mass or HI mass limit of the dwarf galaxies divided by an estimate of the dynamical mass of the galaxy within r$_h$ (log scale) plotted against the distance from the Local Group surface.  Detections are solid diamonds and non-detections are open diamonds.  The lines through the symbols represent the range of possible Local Group surfaces by varying the mass of the MW and M31. We exclude Sag dSph and Bootes III from this plot given their disrupted nature.}
\label{hitotal}
\end{figure*}

The virial radius for M31 from the model in \S\ref{sect:lgsurf} is 266 kpc.   Using this radius as a guide, 27 dwarf galaxies are undetected with limits $\lesssim10^5$ \Msun.  These limits do not vary as much as the Milky Way limits due to the similar distance of these satellites. This count includes M32, AndXXV, and AndXIX that have limits slightly above $10^5$ \msun~due to some confusion with emission from M31 or the Milky Way. The non-detected dwarf galaxies beyond 266 kpc that can be considered satellites of M31 are: AndVI at 268 kpc, AndXXII at 276 kpc, AndXVI at 319 kpc, Perseus I at 348 kpc, AndXXVIII at 365 kpc, and AndXVIII which is at a distant 457 kpc.  
Cetus is a non-detection at 678 kpc from M31 and is in the same category as Tucana, that its large distance makes a direct association with either galaxy uncertain.  KKR25, at close to 2 Mpc from either galaxy, is a non-detection that clearly cannot be associated with either galaxy. The HI detections within the 266 kpc radius for M31 are its dwarf elliptical satellites (NGC205 and NGC185) and IC 10.  The dwarf ellipticals have smaller amounts of HI relative to their luminosity and dynamical mass compared to other dwarf galaxies (Figures~\ref{histar} and \ref{hitotal}; see also \citet{geha06}), and IC 10 is at 252 kpc from M31 and has irregular/disturbed HI, similar to LGS3 at 267 kpc \citep{ashley14,hunter12}.  This may indicate the HI gas of these galaxies is beginning to be removed.  As can be seen from Figure~\ref{himwm31}, the outer range of possible M31 virial radii (279 kpc) considered for our Local Group model results in 29 dwarf galaxies within this radius without gas, and 4 with gas.

Figure~\ref{hilg} shows the HI measurements in the context of the dwarf galaxy's distance from a Local Group surface, or an approximate virial radius of the Local Group. As outlined in \S\ref{sect:lgsurf}, this surface was defined using the MW and M31 mass and concentration parameters, and variation in this surface was calculated by varying the mass of the Milky Way and M31.  The range of distances relative to the variation in the Local Group surface is shown by the line through each symbol, with our fiducial model represented by the position of the symbol and noted in Table~2.  Immediately apparent is, as with the MW and M31, the non-detections are largely located within the Local Group surface (0 on the plot), while those with detected HI are beyond this Local Group surface. There are 6 detected exceptions within the surface, 3 of which are the most optically luminous dwarf galaxies in the sample (LMC, SMC, NGC 205) and two others in the top 11 most luminous (IC10 and NGC185).  These detection exceptions are the same as for the MW and M31, with the addition of LGS3, which was right at the edge of M31's virial radius and is very close to the edge of the LG surface.  The non-detections beyond the LG surface are AndXVIII, Perseus I, Eridanus II, Leo I, and Leo II.  Cetus, Tucana and the very distant KKR25 remain odd non-detections at large radii similar to the MW and M31.
As can be seen, some dwarf galaxies move inside or outside of the Local Group surface with higher or lower Milky Way and M31 masses.  With a 25\% lower Milky Way mass, but higher M31 mass to obtain the same total Local Group mass, 7 (instead of 8) non-detections would be beyond the surface; and with a 25\% higher Milky Way mass and correspondingly lower M31 mass, 9 non-detections would be beyond the Local Group surface.  If the Local Group is more massive than the $3.2 \times 10^{12}$ \Msun~used here, the surface would easily encompass the 4 non-detections within 50 kpc of the surface.  Using our fiducial model as the primary comparison point, we find the Local Group surface has 8 non-detections beyond the surface and the MW and M31 have a total of 13 non-detections beyond their virial radii.  This value of 13 does not change with the changes in the masses of the MW and M31.  The Local Group surface therefore encompasses more of the galaxies without gas.

The completeness of the optical surveys in finding dwarf galaxies at various radii should be considered when assessing the results for both the virial radii of the Milky Way and M31 and the Local Group surface.  Most of the dwarf galaxies were discovered in SDSS, DES, PanSTARRS, or PANDAS \citep[e.g.][]{willman05b, belokurov07, bechtol15, koposov15, laevens15, martin06}, and though the completeness has been partially assessed for some of these surveys \citep{walsh09, newton18}, none of these surveys cover the complete sky.   If one adopts the all-sky survey of \cite{whiting07}, that is noted to be 77\% complete for objects brighter than 25 mag arcsec$^{-2}$ in R, all of the galaxies within 275 kpc of the Milky Way or Andromeda are devoid of gas.  Likewise, galaxies within the Local Group surface, or close to the surface used here (e.g., Leo I and Leo II) are devoid of gas.  Beyond these distances things are less clear, with 4 galaxies detected in HI and 3 galaxies with HI mass limits.   Though the \cite{whiting07} sample is very limited in number of dwarf galaxies, it may suggest additional faint galaxies without gas will be detected at larger radii.

Figures~\ref{histar} - \ref{hitotal} examine if the HI limits for the dwarf galaxies are significant when scaled by their luminosity or dynamical mass.  In another words, one might expect the amount of gas in a dwarf galaxy to correlate with the stellar mass or total mass and therefore the limits should be lower than the detected galaxies when normalized. The V-band luminosity can be considered as a proxy for the stellar mass, consistent with the stellar mass-to-light ratio of 1 used in M12.  While the stellar mass to light ratio will vary for different dwarf galaxies, we do not address that here as these estimates can have many complications and do not span more than an order of magnitude.  Figure~\ref{histar} shows this scaling of the HI content of the dwarf galaxies by V-band luminosity (see Table~\ref{tab:other}).  The scaled HI limits remain significant, however there are a number of undetected dwarf galaxies around \Mhi/L$_{\rm V}=1$.  The median value of \Mhi/L$_{\rm V}$ for the HI detected dwarf galaxies is 0.93 (blue line on Figure~\ref{histar}), and the average value is 1.2.  The undetected dwarf galaxies near these values are dominated by ultra-faint dwarf galaxies with very small stellar populations and some Andromeda dwarf galaxies that are faint, but also influenced by slightly higher HI mass limits.  All of the undetected Milky Way dwarf galaxies with \Mhi/L$_{\rm V}\sim1$ have L$_{\rm V} < 10^4$ L$_\odot$.  Leo T is the best HI detected comparison point for these galaxies and does have a \Mhi/L$_{\rm V}$ higher than most of these galaxies. Stronger limits on their HI masses would be useful, but the dwarf galaxies near the 1 line have ancient stellar populations and are highly unlikely to have gas that can support star formation.

The dynamical mass estimates for the dwarf galaxies were calculated within the half light radius (r$_{\rm h}$) using the same method adopted by M12 in most cases.  This involves using the measured stellar velocity dispersions and r$_{\rm h}$ values and the equation from \cite{walker09}, $M_{dyn} (\le r_h)$ (\Msun) $= 580 r_h (\rm pc) ~\sigma_\star (\rm km/s)^2$.  When values of r$_{\rm h}$ and $\sigma_\star$ were not in M12, they were added from S19 or the literature, and upper limits were used when necessary (see Table~\ref{tab:other}).   Many of the dwarf galaxies with gas are more distant and do not have a measured stellar velocity dispersion.  For these galaxies, we adopt the dispersion calculated from the FWHM of the HI line and calculate the total mass within r$_{\rm h}$ using 
$M_{dyn} (\le r_h) = 3 r_h \sigma_{gas}^2 / G$, which can be written $M_{dyn} (\le r_h)$ (\Msun) $=  698 r_h (\rm pc) ~\sigma_{gas} (\rm km/s)^2$.  The use of the gas dispersion will result in a larger dynamical mass than if the internal stellar velocity dispersion was used.  These masses are not designed to be representative of the true total mass of the dwarf galaxies, but are used throughout for consistency.  
Figure~\ref{hitotal} shows the results of scaling the HI mass limits and HI masses by this dynamical mass estimate.  It shows that the HI limits are significant when scaled by the total mass interior to r$_{\rm h}$, since the vast majority of the detected dwarf galaxies lie above the non-detections.  The non-detections with higher \Mhi/M$_{\rm dyn}$~values are largely faint Andromeda dwarf galaxies that do not have as deep of HI limits. We exclude the Sagittarius dwarf spheroidal galaxy and Bootes III from this plot since their stellar components are clearly not in equilibrium.  For the detections, besides the dwarf ellipticals, Phoenix is somewhat of an outlier in that it is lower than other detections in both this and the \Mhi/L$_{\rm V}$ plot.  Phoenix is at a velocity of -13 \kms~and confused with Galactic emission.  This location on the plot may indicate more of the gas mixed in with Galactic should be associated with the dwarf galaxy, that it is in the process of being stripped, or possibly that the gas is not actually associated with the dwarf.

\section{Discussion}

In this section we discuss the results in the context of the likely methods of gas removal from Local Group dwarf galaxies.  We separate the discussion into the effect of a Milky Way (or M31) gaseous halo medium (\S5.1), and the potential role of a diffuse Local Group gaseous medium in removing gas from dwarf galaxies (\S5.2).  We briefly discuss other methods to potentially remove gas from dwarf galaxies in \S5.3.

\subsection{Stripping by a Spiral Galaxy Halo Medium}

The halo media of the Milky Way and M31 have been thought to play a dominant role in setting the relationship between the gas content of dwarf galaxies and their distance from these large spirals \citep[GP09,][]{spekkens14, blitz00}.  Numerous dwarf galaxies without gas have now been added within the virial radii of the MW and M31, and this provides additional support for a diffuse halo medium playing an important role in quenching the dwarf galaxies.  The few exceptions of HI detections within the virial radii of these galaxies were already known and consistent with being more massive galaxies that can more easily retain their gaseous reservoirs \citep[e.g.][]{simpson18, garrison19}.  These galaxies also often have gas that is disturbed in nature and may be in the process of being removed.  The below discussion focuses on the halo medium of the Milky Way, however we note that M31 has a halo medium that is likely to be similarly effective at stripping dwarf galaxies \citep{lehner20}.

 We can freshly examine the required gas density to ram pressure strip Milky Way dwarf galaxies for those that have proper motion measurements from \textit{Gaia} \citep[see Table~\ref{tab:orbit};][]{fritz18,helmi18,pace19,simon18}.
The criteria to ram pressure strip a galaxy of its gas was originally derived by \cite{gunn72}.   
The equation that describes the complete stripping of a homogeneous medium can be written,
\begin{equation}
n_{\rm halo} \sim \frac{\sigma^{2}\,n_{\rm gas}}{v_{\rm sat}^{2}} \;
\label{eq:halo}
\end{equation}
where $n_{\rm halo}$ is the ambient gas number density in the halo in cm$^{-3}$, $\sigma$ is the stellar velocity dispersion of the dwarf, $v_{\rm sat}$ is
the relative motion of the dwarf through the medium, and $n_{\rm gas}$ is the average gas density of the dwarf in the inner regions.  Some comparisons with numerical simulations have found this equation tends to under-estimate the halo density required for stripping.   \cite{gatto13} completed 2D simulations of the stripping of classical dwarf galaxies and found the equation gives values approximately a factor of 5 too low.   \cite{salem15} also find a somewhat lower value for the analytical calculation for the stripping effect on the LMC, but the value is within the errors found with the numerical simulation.  We leave a more detailed examination of the required halo density to future simulation work, but account for this potential large systematic by including large error bars on our calculations.

 The internal velocity dispersions ($\sigma$) for the Milky Way dwarf galaxies with proper motions are measured in most cases, albeit with large error bars.  For Hydra II and Triangulum II the measured upper limits are used (see Table~\ref{tab:other}).  A dwarf galaxy's dispersion may be similar to its value at earlier times if the dwarf has not been significantly affected by tidal forces and has not grown substantially. 
 
 The correct gas density in the inner regions of the dwarf galaxies at the time of stripping is a difficult value to pinpoint.   The central densities from model fits to the HI gas profiles depend on the dark matter profiles used and tend to be very high \citep{emerick16, faerman13}.  On the other hand, star formation can be triggered when gas is compressed as it interacts with a diffuse medium, which can result in a decrease in the gas density.  This decrease in the central regions is seen in several galaxies in our sample \citep[e.g., IC~1613 and Sextans A;][]{hunter12}.   An estimate for the central gas density can be obtained using the current gas distribution in Leo T.  \cite{adams18} find a peak column density in the central regions of $4.6 \times 10^{20}$ \cm~in their 117 $\times$ 32 parsec beam and they find an HI extent of 400 pc for the dwarf.  This then results in an average gas density of 0.37 \ccm, which is within the range of values used in GP09 and similar to the typical values found in models within the central 100 pc \citep{emerick16}.  This value is also consistent with what has been found for some larger dwarf galaxies, with central density measurements more typically approaching $10^{21}$ \cm~and extents in the few kpc \citep{hunter12}.

\begin{figure}
\hspace{-6mm} 
\includegraphics[scale=0.53,angle=0]{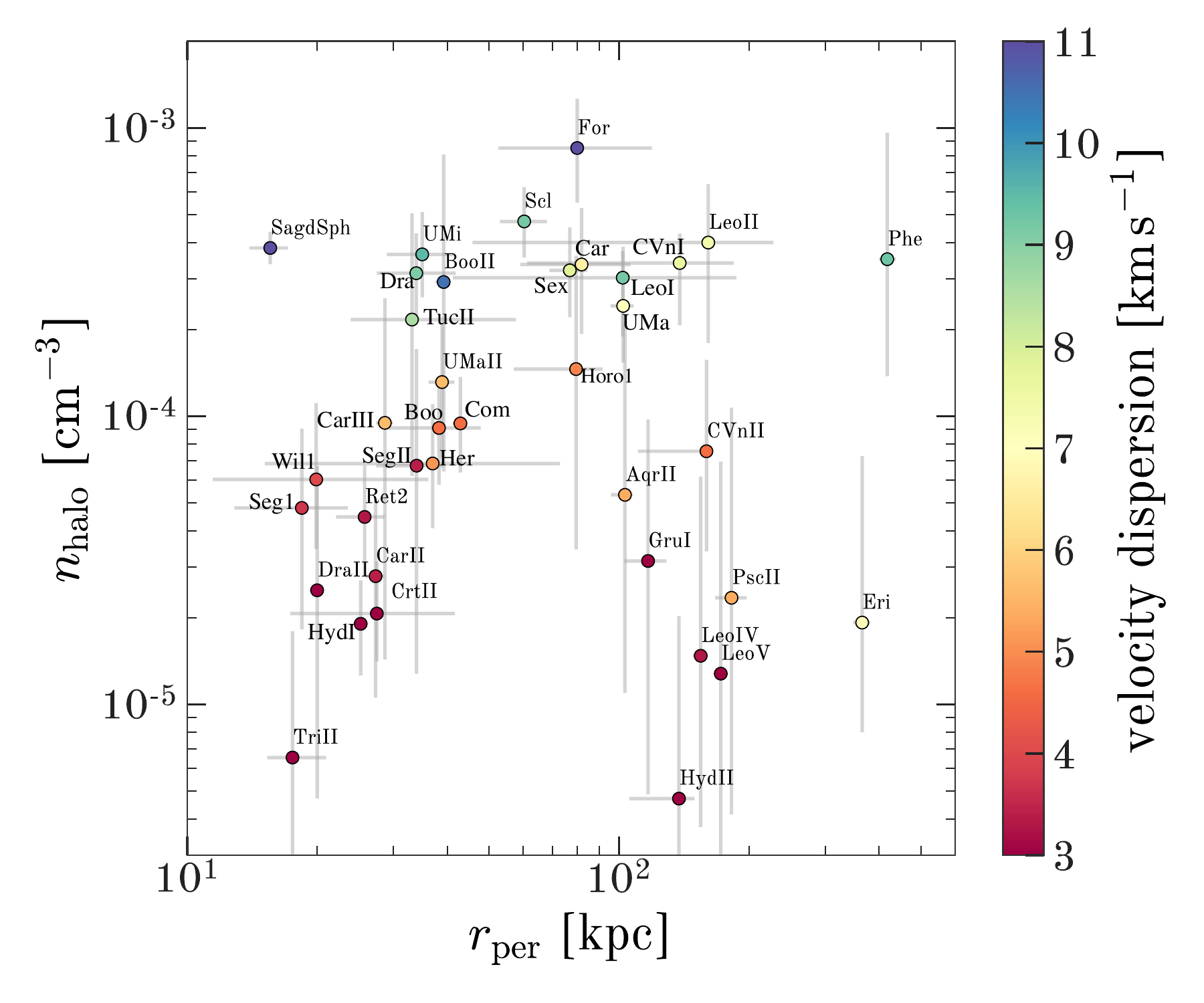}
\vspace{-6mm}
\caption{\small The minimum halo density ($n_{\rm halo}$) required to completely strip the dwarf galaxies with proper motion measurements when at orbital perigalacticon (r$_{per}$).  The median $n_{\rm halo}$ is plotted given the errors on the quantities in Equation~\ref{eq:halo}, with the upper and lower error bars representing the 16th and 84th percentiles. See text and Table~\ref{tab:orbit} for the details.  The velocity dispersions of the dwarf galaxies are color-coded here to provide an estimate of which galaxies are easiest to strip based on this variable. Tucana III is not included on this plot as it has a pericenter of only 2 kpc and requires only $3.5 \times 10^{-7}$ \ccm~to be stripped. We do include the Sag dSph, though its dispersion is confused by its tidal disruption.  Phoenix (Phe, the outlier to the upper right) is the only galaxy on this plot that has gas. }
\label{fig:peri}
\end{figure}

Since a dwarf galaxy will be moving at the highest velocity and most likely through the densest halo medium at perigalacticon, we can assume this would be the point the galaxy would be subject to maximal ram pressure stripping.
We therefore compute the minimum halo gas density, $n_{\rm halo}$,  required to strip the dwarf at perigalacticon using Equation~\ref{eq:halo}, assuming all of the stripping happens at this orbital phase for each dwarf.
We note that this is strictly an upper limit on the minimum value, as it does not account for the steady ram pressure stripping from the extended gaseous halo during its orbit.  We determine the velocity at perigalacticon for those dwarf galaxies with proper motion estimates using the Milky Way model and orbit calculations described in  \S\ref{sect:mwmodel}.
  We compute estimates of $n_{\rm halo}$ and uncertainties on this quantity by Monte Carlo sampling from the error distributions over the velocity dispersion measurements (as compiled from other work) and pericentric velocity for each dwarf galaxy. 
The error samples in pericentric velocity are computed by sampling from the observed error distribution over distance, proper motion, and radial velocity and numerically integrating the orbits of each dwarf galaxy for the range of Milky Way masses describe in \S\ref{sect:mwmodel}.
We then compute the median (over samples) value of $n_{\rm halo}$ (plotted as points in Figure~\ref{fig:peri}) and report upper and lower error bars by computing the 16th and 84th percentiles of our samples over $n_{\rm halo}$. 
We do not account for uncertainty on the gas density of each dwarf because we use a fixed value of $n_{\rm gas}$.   The orbital values and resulting halo densities are tabulated in Table~\ref{tab:orbit}.

We note that whether all of the dwarf galaxies with measured proper motions are bound remains dependent on the mass of the Milky Way and their potential association with the Magellanic Clouds. 
In particular, Hyd I, Hor I, Car II, and Car III are likely Magellanic satellites \citep{patel20, simon18} that may be coming in for the first time.  In any case, these satellites do not stand out in pericenter or halo density compared to other satellites. 

The results of the calculation of the minimum halo density at perigalaction required to strip the galaxy are shown in Figure~\ref{fig:peri}.  For the majority of the dwarf galaxies, the requisite halo gas densities are between $10^{-5}$ to $5 \times 10^{-4}$ \ccm~out to 200 kpc.  The exceptions are two ultrafaint dwarf galaxies that require $<10^{-5}$ \ccm~to be stripped, and Fornax, which requires $\sim10^{-3}$ \ccm~to be stripped.  Phoenix and Eridanus have perigalactica $>300$ kpc; the former has gas, though may show some signs of initial stripping, and the latter requires densities of only $2 \times 10^{-5}$ \ccm~to be stripped.  All galaxies with dispersions $< 6$~\kms~require halo densities $<1.5 \times 10^{-4}$ \ccm~to be stripped at perigalacticon.
Though direct measures of the density of the Milky Way's halo medium are not available, this type of halo density is easily attainable within 50 kpc with the observational constraints that exist \citep{salem15, sembach03, stanimirovic06, putman11, hsu11, miller15, Nidever:2019}.  At larger radii, we do not have solid observational constraints on the halo density of the Milky Way, but absorption line observations indicate a halo medium is prevalent out to the virial radius of galaxies \citep{werk14,tumlinson13,liang14}.  The halo densities required for stripping derived here are overall consistent with the values derived from previous related work for some classical dwarfs given the differences in orbital information and central gas densities \citep[e.g., GP09,][]{gatto13}.

\begin{figure}
\includegraphics[scale=0.63,angle=0]{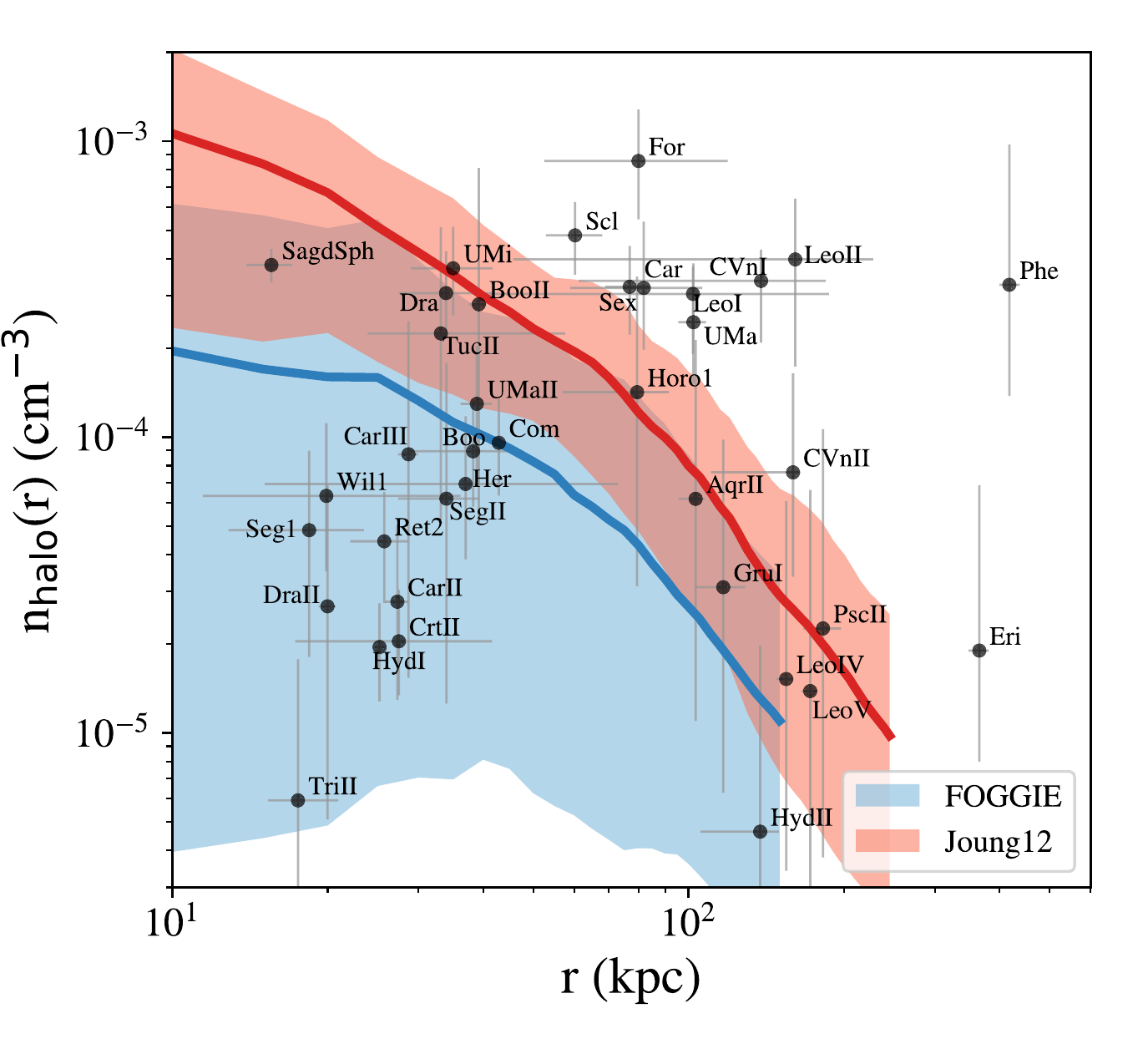}
\vspace{-5mm}
\caption{\small The volume-weighted halo density profile with radius, $n_{\rm halo}(r)$, for two spiral galaxy simulations from FOGGIE \citep[][blue]{peeples19, zheng20} and from \cite{joung12} (red). The cold gas with $T\lesssim10^{4.2}$~K is excluded from this plot to remove the dense disk ISM and satellite gas. The profiles are overlaid with the minimum gas density required to strip a given dwarf galaxy at perigalacticon from Figure~\ref{fig:peri}. The FOGGIE Milky-Way analog (blue) has a mass of $4.9\times10^{11}~\msun$ at $z=0.1$ within $R_{\rm 200}=161.5$ kpc, while the \cite{joung12} simulation has a total mass of $1.4\times10^{12}~\msun$ at $z=0$ within 250 kpc. For each simulation, the solid line denotes the volume-weighted median value and the colored bands show the 5th and 95th volume-weighted percentiles.  }
\label{fig:simhalo}
\end{figure}

The derived halo densities can be compared to the halo density with radius values found in simulations as shown in Figure~\ref{fig:simhalo}.  There are numerous simulations from which the halo gas density can be compared, though most do not plot the data in terms of gas atoms per cubic cm verses radius \citep{putman12,salem16,ford16, vandervoort19}.  When the data are made available in this form, the mean is often adopted, which is biased by clumpy higher density values and results in higher halo densities at a given radius \citep{kaufmann09,nuza14}.  Indeed, \cite{simons20} find the effect of ram pressure stripping on a satellite galaxy can be highly stochastic owing to the broad dynamic range in density and velocity of the CGM medium.  Therefore, in Figure \ref{fig:simhalo} we choose to show the volume-weighted median, and include the 5th and 95th percentiles\footnote{To calculate the volume-weighted median density and percentiles, we first rank all the CGM cells in ascending order according to their volume densities. Then we sort the cells' volumes using the rank and calculate the cumulative distribution function (CDF) of the sorted cells. The volume-weighted median, 5th, and 95th percentiles are the halo densities at which the CDF is equal to 0.5, 0.05, and 0.95, respectively.} to highlight the broad range of densities the dwarf galaxies may experience. 
The simulations shown are two ENZO cosmological simulations of Milky Way analogs: \cite{joung12} that has a total mass of $1.4\times10^{12}~\msun$ at $z=0$ within 250 kpc (red), and the Tempest halo from the FOGGIE simulation \citep{peeples19, zheng20} that has a mass of $4.9\times10^{11}~\msun$ at $z=0.1$ within $R_{\rm 200}=161.5$ kpc (blue).  The Tempest halo from FOGGIE shows lower median values and a broader density spread than the other halo. This difference is likely due to Tempest's lower halo mass and FOGGIE's new refinement scheme that uniformly resolves the CGM of the simulated galaxy in more detail.

We focus on the \cite{joung12} simulation (red) for most of the direct comparison to the dwarf galaxies given the Tempest halo from FOGGIE is not as massive as we expect the Milky Way to be.  The FOGGIE simulation has higher resolution, and though the spread of properties increases with resolution, the average halo densities remain similar \citep{corlies18, peeples19}.   
The halo densities from \cite{joung12}'s simulation are consistent with the majority of the dwarf galaxies being completely stripped as they move through perigalacticon. This is consistent with the simulations of \cite{fillingham19} that find the majority of the galaxies were quenched rapidly on infall.  As mentioned previously, Phoenix still has gas, so the fact that it has not passed through a halo medium dense enough to strip it makes sense.  There is a group of dwarf galaxies with perigalactica clustered roughly around 100 kpc that are not as easily stripped by this simulated $z=0$ halo medium.  Most of these are more massive with velocity dispersions of 6.6 - 11.7 \kms, and Fornax and Carina have longer quenching timescales and more circular orbits consistent with them being less likely to have instanteous stripping \citep{fritz18, fillingham19}.  Some of the dwarf galaxies may have been stripped as they passed through an overdense region of halo gas or star formation triggered by the compression of the gas may have helped to loosen the gas for stripping \citep{simons20,wright19,fillingham16}.
Halo densities required for stripping are also likely to be more easily reached at earlier times, as cosmological simulations indicate the gaseous surroundings were denser and colder in the past \citep{fernandez12,rahmati16}.  Early stripping is consistent with the fact that these galaxies ceased their star formation at early times.

Given the low densities required for stripping the lowest mass dwarf galaxies at perigalacticon, we checked the required densities to strip the dwarf galaxies at apogalacticon (see Table~\ref{tab:orbit}).  This indicates if these dwarf galaxies could have been stripped early in their orbit, or even on entry into the Local Group (see next section).  The orbital values at apogalacticon are more uncertain, but generally gas densities $> 10^{-4}$~\ccm~are required; densities that are unlikely to be reached at large radii.  The exceptions are a group of small galaxies at apogalactica $>1$ Mpc that have required densities of only $\sim10^{-5}$~\ccm.  The level of stripping that a dwarf galaxy experiences throughout its orbit will depend on star formation loosening the gas and 
the orbital direction of the dwarf galaxy relative to the movement of the halo medium.
There are clear indications that the halo medium of a galaxy rotates \citep{defelippis20,hodges16,martin19,oppenheimer18,simons20}, and this would potentially make stripping more (or less) effective via the addition of another velocity component between the gaseous medium and the dwarf galaxy.

\subsection{Stripping by a Local Group Medium}
\label{lg}

As shown in Figure~\ref{hilg}, the Local Group dwarf galaxies with and without gas not only follow a relationship with distance from the Milky Way and Andromeda, but also with distance from a Local Group virial radius or surface.  The relationship with the Local Group is potentially better in the sense that 8 non-detections are beyond the Local Group surface, while 13 are beyond the virial radii of the MW or M31.  Dwarf galaxies close to the Local Group surface line can change to be beyond or within this line with reasonable choices of the model parameters (see \S\ref{sect:lgsurf}), but the Local Group surface was defined using the MW and M31 mass models that define their virial radii for Figure~\ref{himwm31}.  We find at most 1 galaxy moves inside or outside the LG surface with variations in the mass of the MW and M31, while the number inside or outside the virial radii of these galaxies remains the same.  It should also be noted that timing argument masses for the Local Group are generally higher than the $3.2 \times 10^{12}$ \Msun~used here, and any increase in the total mass of the Local Group would easily bring the 4 non-detections close to the Local Group surface within that surface \citep{VanDerMarel:2012, li08}.  The virial radii of the fiducial MW and M31 used here (224 kpc and 266 kpc, respectively), are such that the halos of the two large spirals are within 289 kpc of touching each other.  On one hand, this makes the relationship with Local Group surface not surprising; on the other hand, this indicates the importance of considering the effect of a group medium in stripping gas from dwarf galaxies.

The Local Group is a fairly low mass group and there are limited predictions for the properties of an intragroup medium at this mass scale.  One comparison point is the Local Group simulation of \cite{nuza14} where the two galaxies are embedded in an elongated distribution of hot ($>10^5$ K) gas, with $\sim20\%$ below this temperature range.  The mean gas density declines below $10^{-5}$ \ccm~beyond 300 kpc from the Milky Way in most directions, with the distinct exception of the direction towards M31 where the density is $4 \times 10^{-5}$ \ccm~at its lowest point and then increases again.   It is unlikely the orbits of the dwarf galaxies have changed greatly with redshift \citep{wetzel11}, but unfortunately the orbital information derived from the observations is insufficient to indicate how many have passed through the denser region between the Milky Way and M31 in the past.  The typical Local Group densities in simulations are unlikely to completely strip most dwarf galaxies, but it will loosen the gas \citep{emerick16}, and there are a few possible exceptions for complete stripping in our sample.  Hydra II, Leo IV, Leo V, Pisces II, and Eridanus II are fast-moving small dwarf galaxies that would only require densities of $\sim10^{-5}$ \ccm~to be completely stripped along most of their orbit (see Table~\ref{tab:orbit}).  

Though the Local Group medium may not be able to completely strip most dwarf galaxies, it is likely to be able to strip away any diffuse halo medium, thus leading to a gradual starvation of the galaxy.  This starvation of gas that could have accreted to feed star formation is found for satellites in massive halos in the EAGLE simulation \citep{vandevoort17}.  In particular, the more massive the host halo, the greater the starvation of the satellites. \cite{marasco16} find the same thing for removal of HI from galaxies, that more massive host halos are more effective with the removal, although this is for galaxies with stellar masses LMC and greater.  The group environment may have also led to more satellite-satellite interactions that helped to loosen the gas \citep{marasco16,pearson18}.  Furthermore, any dwarf galaxies that came in with the Magellanic System are likely to have been pre-processed at some level \citep{patel20}.  In summary, the simulation results indicate the halo of the Local Group would be expected to have a larger quenching role than an individual Milky Way mass galaxy \citep[see also][]{garrison19}. The group environment therefore may help to explain why some observed isolated Milky Way analogs do not show the same large fraction of quenched satellites \citep{geha17}.  

Direct evidence for a Local Group medium, or diffuse hot gas filling the volume beyond the virial radii of the Milky Way and Andromeda, is not available, but there are some indirect suggestions that this medium exists.   The HI structure of some of the galaxies close to the edge of the Local Group, but beyond the virial radius of the Milky Way or M31, have been claimed to be affected by ram pressure stripping by a diffuse gaseous medium (e.g., Leo T, Pegasus, and possibly Phoenix \citep{adams18,mcconnachie07, young07,stgermain99}).
In addition, the HI structure of the Magellanic System can be more easily explained with the presence of a Local Group medium.  The Clouds are modeled to be coming into the Milky Way halo for the first time and are currently at 50-60 kpc \citep{besla10};  yet a huge amount of ionized and neutral gas trails behind the Clouds, with less on the leading side of the Clouds \citep{putman98,fox14,putman03b}.  The gas is thought to have been initially loosened from the interaction of the Clouds with each other, but a more prolonged passage through a gaseous medium would help create the $>100$ kpc massive tail of gas.  Finally, there are detections of ultraviolet absorption lines that remain mysterious in origin (e.g., velocities not easily connected to nearby denser structures and very low derived pressures) that may be related to a Local Group medium \citep{sembach03,richter2017}.  This would be consistent with the UV COS results of \cite{stocke19}, who find a bias for intragroup gas detections in the lower mass, more spiral rich groups.  They only probe groups with mass $> 10^{13.5}$ \Msun, but predict from their results that Local Group mass scale groups will have a 100\% covering fraction of diffuse gas out to a couple of virial radii.

\subsection{Other Methods of Gas Removal}

As mentioned in the introduction, there are several mechanisms besides ram pressure stripping that could play a role in removing gas from dwarf galaxies, including stellar feedback, tidal forces and reionization.  There is not strong evidence that stellar feedback or tidal forces are dominant mechanisms given the distance dependent effect shown here and the strength of the tidal force during the orbit of the dwarf galaxies and lack of stellar tidal features \citep[GP09]{mayer06, mateo08, simpson18, blitz00, emerick16}.  However, given many dwarf galaxies without gas are likely to have lost it in the early universe \citep{weisz14},    
reionization needs to be considered at the lowest mass scales \citep{gnedin00,kang19,rodriguez19}.  The finding of Leo T, a small dwarf galaxy with gas, cast doubt on the effectiveness of reionization, but the lack of additional galaxies with gas at this mass scale is supportive of it \citep[e.g.][]{tollerud18}.  The finding of more small dwarf galaxies without gas beyond the edges of the Local Group will provide support for reionization being an important quenching mechanism.  The proper motion measurements find fewer dwarf galaxies at apocenter than pericenter, and this indicates additional dwarf galaxies are likely to be found at larger distances \citep{fritz18}.  Simulations also expect there to be additional dwarf galaxies in the Local Group beyond the virial radii of the Milky Way and M31 \citep{klypin15,fattahi20}.

Thus far, the gas-less dwarf galaxies at larger distances (e.g., Tucana, Cetus, AndXVIII, Eridanus II) may belong to a population of backsplash galaxies, or galaxies predicted by simulations to have fallen in at early times and subsequently flung out to larger distances \citep{teyssier12,simpson18,fillingham18,blana20}.  In the study of simulated Milky Way analogs by \cite{simpson18}, 41\% of the systems beyond the virial radius and within 1 Mpc are backsplash galaxies, indicating that more gas-less dwarf galaxies may be discovered at large radii that are not necessarily linked to reionization (see also \cite{wetzel14}).  It may be possible to obtain proper motions for gas-less dwarf galaxies that are currently beyond the virial radius in the future (e.g., Eridanus II) to determine if they are likely to have been relatively deep within the Local Group in the past.  KKR25 is a larger (i.e., too big for reionization quenching) dwarf galaxy at large distance for which the quenching mechanism is unknown, and it is joined by two others at distances greater than 2 Mpc \citep{sharina17, karachentsev14}. 
In any case, at early times a combination of quenching mechanisms is likely to have played a role; however reproducing the distance dependency shown here is likely to be difficult without ram pressure stripping.   Future observations of the field galaxy population will help to further distinguish between stripping mechanisms.

\section{Summary}
We have placed limits on the gas content of the dwarf galaxies within 2 Mpc of the Milky Way, and examined the relationship of gas content with distance from the Milky Way and M31 and relative to their location in the Local Group.   The findings can be summarized as follows:
\begin{itemize}

\item HI gas mass limits for the vast majority of the non-detected dwarf galaxies are less than $10^5$~\Msun~($5\sigma$).     
This is less than the HI mass of any of the detected dwarf galaxies. This limit improves to $<10^4$~\Msun~for the dwarf galaxies within the virial radius of the Milky Way. Scaling by stellar mass (i.e., optical luminosity) or total mass (within the stellar component) the limits are consistent with the gas content of the dwarf galaxies being far suppressed from where they were when they formed their stars.

\item The number of dwarf galaxies within 2 Mpc has doubled in the last 10 years, but there is a distinct lack of new dwarf galaxies with gas.  This suggests galaxies like Leo T are either rare or only at distances where current surveys struggle to detect them \citep[see][]{tollerud18,defelippis19}.  Future surveys capable of detecting low mass dwarf galaxies at large distances in gas \citep[e.g. WALLABY;][]{koribalski20} and stars \citep[e.g., LSST;][]{ivesic19} will provide important insight into the dwarf population and quenching mechanisms.

\item There is a clear relationship that those dwarf galaxies with gas are beyond the virial radii of the Milky Way and Andromeda, while non-detections are within these radii. This relationship of gas-content vs. distance is also found when the distance from a Local Group surface, or virial radius, is used. More of the non-detected dwarf galaxies are within the Local Group surface than the virial radii of the MW and M31 (85$\pm1$ vs.~80, respectively), which may indicate a Local Group medium plays a role in the gas stripping. 

\item  The relationship of gas content with distance suggests a distance dependent quenching mechanism such as ram pressure stripping by a diffuse gaseous medium.  Using the proper motions available from \textit{Gaia}, the orbits of 38 dwarf galaxies are calculated and the minimum required densities for stripping at perigalacticon are typically between $10^{-5}$ to $5 \times 10^{-4}$ \ccm.  Compared to the halo densities found in simulations, 80\% of these dwarf galaxies are consistent with being stripped at perigalacticon.  Continuous stripping throughout their orbit is likely to play an important additional role in starving and quenching the dwarf galaxies.  
\end{itemize}

\acknowledgements{The authors acknowledge the GALFA-HI team that enabled much of the data used to exist, useful discussions with Stephanie Tonneson, Andrew Emerick, Juergen Kerp, Kathryn Johnston, Hsiao-Wen Chen, Greg Bryan, Ekta Patel, Tobias Westmeier, Raymond Simons and Molly Peeples, and useful comments from an anonymous referee.  MEP and APW acknowledge KITP UC-Santa Barbara where we partially worked on this, which is supported by the National Science Foundation under Grant No. NSF PHY-1748958.  MEP acknowledges childcare support from Daniel and Elaine Putman during the unique time this paper was submitted.}

\clearpage

\begin{center}
\begin{ThreePartTable}
\begin{TableNotes}
\label{tab:hi}
\item Notes -- 
(1) The data source is either HI4PI \citep{hi4pi}, GALFA-HI DR1 \citep[GALFA1, ][]{peek11}, or GALFA-HI DR2 \citep[GALFA2, ][]{peek18}.
(2) The method indicates if the $1\sigma$ was obtained with a method consistent with the source being resolved (Res.) or unresolved (Unres.) in the data.  If the source does not have a velocity (No V), then the $1\sigma$ is an average from all of the channels without significant Galactic emission.  For HI4PI resolved, the $1\sigma$ values tend to be lower as they are the standard deviation of the flux sum divided by the area of the optical galaxy for 9 regions surrounding the galaxy (see text).
(3) The HI mass is calculated using 5$\sigma$, the noted distance, and a fixed velocity width of 10 \kms.  The errors are from the distance errors shown.
(4) This column notes the references used for the positions, velocities and distances, respectively.
(5) \cite{mcconnachie12}.
(6) \cite{simon19}.
(7) \cite{luque17} for the position.
(8)  The limit is only for the core of this tidally extended galaxy.
(9) \cite{koposov18} for the position.
(10) \cite{torrealba18} for the position.
(11) The limit is calculated with twice the median flux in the region of the galaxy due to strong Galactic contamination (see text).
(12) \cite{drlica16} for the position.
(13) \cite{simon20} for the position and velocity.
(14) Sculptor has been claimed as a detection in previous publications, but the HI clouds are offset from the stellar position of the galaxy and there is abundant surrounding emission at similar velocities.
(15) \cite{homma16} for the position.
(16) \cite{torrealba16b} for the position.
(17) \cite{torrealba16} for the position.
(18) \cite{torrealba19} for position, velocity and distance.
(19) Fornax is at velocities that confuse any associated HI with Galactic emission.  The limit at the stellar position and velocity is given.
(20) This galaxy is unlikely to exist based on observations by \cite{cantu20}.
(21) \cite{homma18} for the position.
(22) Phoenix has confusion with Galactic emission, but the HI cloud is at the position and velocity of the stellar component.
(23) \cite{martin14}. 
(24) Lacerta I is Andromeda XXXI in \cite{mcconnachie12}.
(25) Perseus I is Andromeda XXXIII in \cite{mcconnachie12}.
(26) Cassiopeia III is Andromeda XXXII in \cite{mcconnachie12}.
(27) \cite{bouchard05} found a velocity of 159 \kms, but we do not note that here as this could not be confirmed by \cite{westmeier17} or \cite{koribalski18}. \cite{koribalski18} find emission with a similar flux and velocity width at 36 \kms, so we leave the HI mass.
\end{TableNotes}
\begin{longtable*}{llcccccrc}
\caption{Dwarf Galaxies within 2 Mpc:  Positions and HI Parameters} \\
\hline
Galaxy & RA and DEC & V$_\odot$  & D$_\odot$ & Data$^1$ & Method$^2$ & $1\sigma$  & M$_{\rm HI}^3$  & Refs.$^4$ \\
 & & (km/s) & (kpc) & & & (mK) & (\msun) & \\
\hline
\endfirsthead
\hline
Galaxy & RA and DEC & V$_\odot$  & D$_\odot$ & Data$^1$ & Method$^2$ & $1\sigma$  & M$_{\rm HI}^3$  & Refs.$^4$ \\
 & & (km/s) & (kpc) & & & (mK) & (\msun) & \\
\hline
\endhead
\endfoot
\hline
\insertTableNotes
\endlastfoot
Draco II &  $15^ {h}52^ {m}48^ {s}~ $+$64^ {\circ}33^ {\prime}55^ {\prime \prime}$ &           -343 &          22$^{\pm0.4}$ &  HI4PI &         Unres. &                   17 &                  $<57^{\pm2}$ &     5,6,6 \\    
Segue 1 &  $10^ {h}07^ {m}04^ {s}~ $+$16^ {\circ}04^ {\prime}55^ {\prime \prime}$ &                    209 &                           23$^{\pm2}$ &  GALFA1 &           Res. &                   17 &                  $<47^{\pm8}$ &     5,6,6 \\      
DESJ0225+0304 &  $02^ {h}25^ {m}42^ {s}~ $+$03^ {\circ}04^ {\prime}10^ {\prime \prime}$ &                      - &                           24$^{\pm1}$ &  GALFA2   &            Unres. No V &                   97 &                  $<71^{\pm4}$ &     7,-,7 \\           
Tucana III &  $23^ {h}56^ {m}36^ {s}~ $-$59^ {\circ}36^ {\prime}00^ {\prime \prime}$ &                   -102 &                           25$^{\pm2}$ &  HI4PI &         Unres. &                   20 &                  $<90^{\pm14}$ &     5,6,6 \\   
Sagittarius dSph$^8$ &  $18^ {h}55^ {m}20^ {s}~ $-$30^ {\circ}32^ {\prime}43^ {\prime \prime}$ &                    139 &                           27$^{\pm1}$ &  HI4PI &         Unres. &                   25 &                 $<126^{\pm12}$ &   5,6,6   \\   
Hydrus I &  $02^ {h}29^ {m}33^ {s}~ $-$79^ {\circ}18^ {\prime}32^ {\prime \prime}$ &                     80 &                           28$^{\pm1}$ &  HI4PI &         Unres. &                   23 &                 $<125^{\pm5}$ &      9,6,6 \\                
Carina III &  $07^ {h}38^ {m}31^ {s}~ $-$57^ {\circ}53^ {\prime}59^ {\prime \prime}$ &                    285 &                           28$^{\pm1}$ &  HI4PI &         Unres. &                   18 &                 $<101^{\pm4}$ &     10,6,6  \\  
Triangulum II &  $02^ {h}13^ {m}17^ {s}~ $+$36^ {\circ}10^ {\prime}42^ {\prime \prime}$ &                   -382 &                           28$^{\pm2}$ &  HI4PI &         Unres. &                   19 &                 $<107^{\pm12}$ &     5,6,6 \\ 
Cetus II &  $01^ {h}17^ {m}53^ {s}~ $-$17^ {\circ}25^ {\prime}12^ {\prime \prime}$ &                      - &                           30$^{\pm3}$ &  HI4PI &    Unres. No V &                   20 &                 $<127^{\pm26}$ &     5,-,6 \\       
Reticulum 2 &  $03^ {h}35^ {m}42^ {s}~ $-$54^ {\circ}02^ {\prime}57^ {\prime \prime}$ &                     63 &                           32$^{\pm2}$ &  HI4PI &         Unres. &                   21 &                 $<145^{+14}_{-13}$ &     5,6,6 \\     
Ursa Major II &  $08^ {h}51^ {m}30^ {s}~ $+$63^ {\circ}07^ {\prime}48^ {\prime \prime}$ &                   -117 &                           35$^{\pm2}$ &  HI4PI &           Res. &                  13 &                  $<113^{+13}_{-12}$ &     5,6,6 \\         
Carina II &  $07^ {h}36^ {m}26^ {s}~ $-$57^ {\circ}59^ {\prime}57^ {\prime \prime}$ &                    477 &                           36$^{\pm1}$ &  HI4PI &         Unres. &                   24 &                 $<221^{\pm7}$ &    10,6,6 \\   
Segue II &  $02^ {h}19^ {m}16^ {s}~ $+$20^ {\circ}10^ {\prime}31^ {\prime \prime}$ &                    -40 &                           37$^{\pm3}$ &  GALFA2 &           Res. &                   43 &                 $<306^{\pm50}$ &     5,6,6 \\          
Boötes II &  $13^ {h}58^ {m}00^ {s}~ $+$12^ {\circ}51^ {\prime}00^ {\prime \prime}$ &                   -117 &                           42$^{\pm1}$ &  GALFA2 &           Res. &                   20 &                 $<185^{\pm9}$ &     5,6,6 \\    
Coma Berenices  &  $12^ {h}26^ {m}59^ {s}~ $+$23^ {\circ}54^ {\prime}15^ {\prime \prime}$ &                     98 &                           42$^{\pm2}$ &  GALFA1 &           Res. &                   12 &                 $<105^{\pm8}$ &     5,6,6 \\          
Willman 1$^{11}$ &  $10^ {h}49^ {m}21^ {s}~ $+$51^ {\circ}03^ {\prime}00^ {\prime \prime}$ &                    -14 &                           45$^{\pm10}$ &  HI4PI &         Unres. &                  360 &  $<5.2^{\pm2.3}\times 10^ {3}$ &    5,6,6  \\    
Pictor II &  $06^ {h}44^ {m}43^ {s}~ $-$59^ {\circ}53^ {\prime}49^ {\prime \prime}$ &                      - &                           45$^{+5}_{-4}$ &  HI4PI &    Unres. No V &                   19 &                 $<277^{+62}_{-49}$ &   12,-,6   \\         
Boötes III$^8$ &  $13^ {h}57^ {m}12^ {s}~ $+$26^ {\circ}48^ {\prime}00^ {\prime \prime}$ &                    198 &                           47$^{\pm2}$ &  GALFA1 &         Unres. &                   38 &                 $<108^{\pm10}$ & 5,5,5   \\          
Tucana IV &  $00^ {h}02^ {m}55^ {s}~ $-$60^ {\circ}51^ {\prime}00^ {\prime \prime}$ &  16 &                           48$^{\pm4}$ &  HI4PI &    Unres. &                   15 &                 $<238^{\pm40}$ &     13,13,6 \\                
LMC &  $05^ {h}23^ {m}34^ {s}~ $-$69^ {\circ}45^ {\prime}22^ {\prime \prime}$ &                    262 &                           51$^{\pm2}$ &      - &              - &                    - &     $4.6^{\pm0.4}\times 10^ {8}$ &     5,5,5 \\            
Grus II &  $22^ {h}04^ {m}05^ {s}~ $-$46^ {\circ}26^ {\prime}24^ {\prime \prime}$ &                      -110 &                           53$^{\pm5}$ &  HI4PI &    Unres. &                   20 &                 $<398^{\pm75}$ &     13,13,6\\           
Tucana V &  $23^ {h}37^ {m}24^ {s}~ $-$63^ {\circ}16^ {\prime}12^ {\prime \prime}$ &                      -36 &                           55$^{\pm9}$ &  HI4PI &    Unres. &                   17 &                 $<360^{\pm118}$ &     13,13,6\\          
Tucana II &  $22^ {h}51^ {m}55^ {s}~ $-$58^ {\circ}34^ {\prime}08^ {\prime \prime}$ &                   -129 &                           58$^{\pm8}$ &  HI4PI &           Res. &                  9 &                 $<215^{\pm59}$ &     5,6,6 \\               
SMC &  $00^ {h}52^ {m}45^ {s}~ $-$72^ {\circ}49^ {\prime}43^ {\prime \prime}$ &                    146 &                           64$^{\pm4}$ &      - &              - &                    - &     $4.6^{\pm0.5}\times 10^ {8}$ &     5,5,5 \\            
Boötes  &  $14^ {h}00^ {m}06^ {s}~ $+$14^ {\circ}30^ {\prime}00^ {\prime \prime}$ &                     102 &                           66$^{\pm2}$ &  GALFA2 &           Res. &                   25 &                 $<564^{\pm34}$ &     5,6,6 \\     
Sagittarius II &  $19^ {h}52^ {m}40^ {s}~ $-$22^ {\circ}04^ {\prime}05^ {\prime \prime}$ &                      - &                           70$^{\pm2}$ &  HI4PI &    Unres. No V &                   20 &                 $<696^{\pm46}$ &     5,-,6 \\        
Ursa Minor  &  $15^ {h}09^ {m}08^ {s}~ $+$67^ {\circ}13^ {\prime}21^ {\prime \prime}$ &                   -247 &                           76$^{\pm4}$ &  HI4PI &         Unres. &                   15 &                 $<601^{\pm63}$ &     5,6,6 \\             
Horologium II &  $03^ {h}16^ {m}32^ {s}~ $-$50^ {\circ}01^ {\prime}05^ {\prime \prime}$ &                      - &                           78$^{+8}_{-7}$ &  HI4PI &    Unres. No V &                   18 &                 $<775^{+159}_{-139}$ &     5,-,6 \\       
Draco  &  $17^ {h}20^ {m}12^ {s}~ $+$57^ {\circ}54^ {\prime}55^ {\prime \prime}$ &                   -291 &                           82$^{\pm6}$ &  HI4PI &           Res. &                 12  &                 $<592^{\pm87}$ &     5,6,6 \\      
Phoenix 2 &  $23^ {h}39^ {m}59^ {s}~ $-$54^ {\circ}24^ {\prime}22^ {\prime \prime}$ &                      - &                           84$^{\pm4}$ &  HI4PI &    Unres. No V &                   17 &                 $<855^{\pm81}$ &     5,-,6 \\          
Sculptor$^{14}$  &  $01^ {h}00^ {m}09^ {s}~ $-$33^ {\circ}42^ {\prime}33^ {\prime \prime}$ &                    111 &                           86$^{\pm5}$ &  HI4PI &         Unres. &                   62 &  $<3.2^{\pm0.4}\times 10^ {3}$ &  5,6,6   \\           
Horologium 1 &  $02^ {h}55^ {m}32^ {s}~ $-$54^ {\circ}07^ {\prime}08^ {\prime \prime}$ &                    113 &                           87$^{+13}_{-11}$ &  HI4PI &         Unres. &                   17 &                 $<917^{+274}_{-232}$ &     5,6,6 \\   
Eridanus 3 &  $02^ {h}22^ {m}46^ {s}~ $-$52^ {\circ}17^ {\prime}01^ {\prime \prime}$ &                      - &                           87$^{\pm8}$ &  HI4PI &    Unres. No V &                   20 &  $<1.1^{\pm0.2}\times 10^ {3}$ &     5,-,5 \\            
Virgo I &  $12^ {h}00^ {m}10^ {s}~ $+$00^ {\circ}40^ {\prime}48^ {\prime \prime}$ &                      - &                           87$^{+13}_{-8}$ &    GALFA2 &            Unres. No V &                   82 &                 $<810^{+242}_{-149}$ &   15,-,6  \\      
Reticulum III &  $03^ {h}45^ {m}26^ {s}~ $-$60^ {\circ}27^ {\prime}00^ {\prime \prime}$ &                      - &                           92$^{\pm13}$ &  HI4PI &    Unres. No V &                   17 &  $<1.0^{\pm0.3}\times 10^ {3}$ &     5,-,6 \\     
Sextans  &  $10^ {h}13^ {m}03^ {s}~ $-$01^ {\circ}36^ {\prime}53^ {\prime \prime}$ &                    224 &                           95$^{\pm3}$ &  HI4PI &           Res. &              8 &                 $<479^{\pm30}$ &     5,6,6 \\     
Ursa Major I &  $10^ {h}34^ {m}53^ {s}~ $+$51^ {\circ}55^ {\prime}12^ {\prime \prime}$ &                    -55 &                           97$^{\pm6}$ &  HI4PI &           Res. &               103 &  $<6.9^{\pm0.8}\times 10^ {3}$ &     5,6,6 \\          
Indus I &  $21^ {h}08^ {m}49^ {s}~ $-$51^ {\circ}09^ {\prime}56^ {\prime \prime}$ &                      - &                          100$^{\pm9}$ &  HI4PI &    Unres. No V &                   20 &  $<1.4^{\pm0.3}\times 10^ {3}$ &     5,-,5 \\            
Carina  &  $06^ {h}41^ {m}37^ {s}~ $-$50^ {\circ}57^ {\prime}58^ {\prime \prime}$ &                    223 &                          106$^{\pm5}$ &  HI4PI &         Unres. &                   14 &  $<1.1^{\pm0.1}\times 10^ {3}$ &     5,6,6 \\        
Aquarius II &  $22^ {h}33^ {m}56^ {s}~ $-$09^ {\circ}19^ {\prime}39^ {\prime \prime}$ &                    -71 &                          108$^{\pm3}$ &  HI4PI &         Unres. &                   21 &  $<1.8^{\pm0.1}\times 10^ {3}$ &    16,6,6  \\         
Crater II &  $11^ {h}49^ {m}14^ {s}~ $-$18^ {\circ}24^ {\prime}47^ {\prime \prime}$ &                     88 &                          118$^{\pm1}$ &  HI4PI &           Res. &                 31 &  $<3.0^{\pm0.06}\times 10^ {3}$ &      17,6,6 \\            
Grus I &  $22^ {h}56^ {m}42^ {s}~ $-$50^ {\circ}09^ {\prime}48^ {\prime \prime}$ &                   -141 &                          120$^{+12}_{-11}$ &  HI4PI &         Unres. &                   21 &  $<2.2^{\pm0.4}\times 10^ {3}$ &     5,6,6 \\  
Pictoris 1 &  $04^ {h}43^ {m}47^ {s}~ $-$50^ {\circ}16^ {\prime}59^ {\prime \prime}$ &                      - &                          126$^{+19}_{-16}$ &  HI4PI &    Unres. No V &                   20 &  $<2.3^{+0.7}_{-0.6}\times 10^ {3}$ &     5,-,6 \\   
Antlia 2 &  $09^ {h}35^ {m}33^ {s}~ $-$36^ {\circ}46^ {\prime}02^ {\prime \prime}$ & 291 & 132$^{\pm6}$ & HI4PI  & Res. & 0.9 & $<111^{\pm10}$ & 18,18,18 \\
Hercules  &  $16^ {h}31^ {m}02^ {s}~ $+$12^ {\circ}47^ {\prime}30^ {\prime \prime}$ &                     45 &                          132$^{\pm6}$ &  HI4PI &         Unres. &                   12 &  $<1.5^{\pm0.1}\times 10^ {3}$ &     5,6,6 \\           
Fornax$^{19}$  &  $02^ {h}39^ {m}59^ {s}~ $-$34^ {\circ}26^ {\prime}57^ {\prime \prime}$ &                     55 &                          139$^{\pm3}$ &  HI4PI &         Unres. &                   25 &  $<3.4^{\pm0.1}\times 10^ {3}$ &  5,6,6    \\             
Hydra II &  $12^ {h}21^ {m}42^ {s}~ $-$31^ {\circ}59^ {\prime}07^ {\prime \prime}$ &                    303 &                          151$^{+8}_{-7}$ &  HI4PI &         Unres. &                   20 &  $<3.2^{\pm0.3}\times 10^ {3}$ &     5,6,6 \\      
Leo IV &  $11^ {h}32^ {m}57^ {s}~ $+$00^ {\circ}32^ {\prime}00^ {\prime \prime}$ &                    132 &                          154$^{\pm5}$ &  GALFA2 &           Res. &                   19 &  $<2.3^{\pm0.1}\times 10^ {3}$ &     5,6,6 \\  
Canes Venatici II &  $12^ {h}57^ {m}10^ {s}~ $+$34^ {\circ}19^ {\prime}15^ {\prime \prime}$ &                   -129 &                          160$^{\pm4}$ &  HI4PI &         Unres. &                   13 &  $<2.4^{\pm0.1}\times 10^ {3}$ &     5,6,6 \\              
Leo V &  $11^ {h}31^ {m}10^ {s}~ $+$02^ {\circ}13^ {\prime}12^ {\prime \prime}$ &                    171 &                          169$^{\pm4}$ &  GALFA2 &         Unres. &                   79 &  $<2.9^{\pm0.1}\times 10^ {3}$ &     5,6,6 \\          
Pisces II &  $22^ {h}58^ {m}31^ {s}~ $+$05^ {\circ}57^ {\prime}09^ {\prime \prime}$ &                   -227 &                          183$^{\pm15}$ &  GALFA1 &         Unres. &                   37 &  $<1.6^{\pm0.3}\times 10^ {3}$ &     5,6,6 \\          
Columba I &  $05^ {h}31^ {m}26^ {s}~ $-$28^ {\circ}01^ {\prime}48^ {\prime \prime}$ &                      - &                          183$^{\pm10}$ &  HI4PI &    Unres. No V &                   22 &  $<5.2^{\pm0.6}\times 10^ {3}$ &     5,-,6 \\        
Pegasus III &  $22^ {h}24^ {m}23^ {s}~ $+$05^ {\circ}25^ {\prime}12^ {\prime \prime}$ &                   -223 &                          205$^{\pm20}$ &  GALFA1 &         Unres. &                   57 &  $<3.1^{\pm0.6}\times 10^ {3}$ &     5,6,6 \\           
Canes Venatici I &  $13^ {h}28^ {m}04^ {s}~ $+$33^ {\circ}33^ {\prime}21^ {\prime \prime}$ &                     31 &                          211$^{\pm6}$ &  HI4PI &         Unres. &                   15 &  $<4.8^{\pm0.3}\times 10^ {3}$ &     5,6,6 \\    
Indus II$^{20}$ &  $20^ {h}38^ {m}53^ {s}~ $-$46^ {\circ}09^ {\prime}36^ {\prime \prime}$ &                      - &                          214$^{\pm16}$ &  HI4PI &    Unres. No V &                   19 &  $<6.2^{\pm0.9}\times 10^ {3}$ &    5,-,6  \\ 
Leo II &  $11^ {h}13^ {m}29^ {s}~ $+$22^ {\circ}09^ {\prime}06^ {\prime \prime}$ &                     78 &                          233$^{\pm14}$ &  HI4PI &         Unres. &                   12 &  $<4.5^{\pm0.5}\times 10^ {3}$ &     5,6,6 \\          
Cetus III &  $02^ {h}05^ {m}19^ {s}~ $-$04^ {\circ}16^ {\prime}12^ {\prime \prime}$ &                      - &                          251$^{+24}_{-11}$ &  HI4PI &    Unres. No V &                   19 &  $<8.4^{+1.6}_{-1.0}\times 10^ {3}$ &    21,-,6  \\             
Leo I &  $10^ {h}08^ {m}28^ {s}~ $+$12^ {\circ}18^ {\prime}23^ {\prime \prime}$ &                    283 &                          254$^{+16}_{-15}$ &  HI4PI &         Unres. &                   22 &  $<1.0^{\pm0.1}\times 10^ {4}$ &     5,6,6 \\         
Eridanus II &  $03^ {h}44^ {m}21^ {s}~ $-$43^ {\circ}32^ {\prime}00^ {\prime \prime}$ &                     76 &                          366$^{\pm17}$ &  HI4PI &         Unres. &                   19 &  $<1.8^{\pm0.2}\times 10^ {4}$ &     5,6,6 \\           
Leo T &  $09^ {h}34^ {m}53^ {s}~ $+$17^ {\circ}03^ {\prime}05^ {\prime \prime}$ &                     38 &                          409$^{+29}_{-27}$ &      - &              - &                    - &     $2.8^{\pm0.4}\times 10^ {5}$ &     5,6,6 \\     
Phoenix$^{22}$  &  $01^ {h}51^ {m}06^ {s}~ $-$44^ {\circ}26^ {\prime}41^ {\prime \prime}$ &                    -13 &                          415$^{\pm19}$ &      - &              - &              - &     $1.2^{\pm0.1}\times 10^ {5}$ &  5,5,5     \\              
NGC 6822 &  $19^ {h}44^ {m}57^ {s}~ $-$14^ {\circ}47^ {\prime}21^ {\prime \prime}$ &                    -55 &                          459$^{\pm17}$ &      - &              - &                    - &     $1.3^{\pm0.1}\times 10^ {8}$ &     5,5,5 \\      
Andromeda XVI &  $00^ {h}59^ {m}30^ {s}~ $+$32^ {\circ}22^ {\prime}36^ {\prime \prime}$ &                   -367 &                          476$^{+42}_{-31}$ &  GALFA2 &         Unres. &                   69 &  $<2.0^{+0.4}_{-0.3}\times 10^ {4}$ &     5,5,5 \\     
Andromeda XXIV &  $01^ {h}18^ {m}30^ {s}~ $+$46^ {\circ}21^ {\prime}58^ {\prime \prime}$ &                   -128 &                          600$^{\pm33}$ &  HI4PI &         Unres. &                   16 &  $<4.0^{\pm0.4}\times 10^ {4}$ &     5,5,5 \\            
NGC 185 &  $00^ {h}38^ {m}58^ {s}~ $+$48^ {\circ}20^ {\prime}15^ {\prime \prime}$ &                   -204 &                          617$^{\pm26}$ &      - &              - &                    - &     $1.1^{\pm0.1}\times 10^ {5}$ &     5,5,5 \\       
Andromeda XV &  $01^ {h}14^ {m}19^ {s}~ $+$38^ {\circ}07^ {\prime}03^ {\prime \prime}$ &                   -323 &                          625$^{+75}_{-35}$ &  HI4PI &         Unres. &                   15 &  $<4.1^{\pm1}\times 10^ {4}$ &   5,5,5 \\        
Andromeda II &  $01^ {h}16^ {m}30^ {s}~ $+$33^ {\circ}25^ {\prime}09^ {\prime \prime}$ &                   -192 &                          652$^{\pm18}$ &  GALFA2 &           Res. &                   20 &  $<4.4^{\pm0.2}\times 10^ {4}$ &     5,5,5 \\   
Andromeda XXVIII &  $22^ {h}32^ {m}41^ {s}~ $+$31^ {\circ}12^ {\prime}58^ {\prime \prime}$ &                   -326 &                          661$^{+152}_{-61}$ &  GALFA1 &         Unres. &                   31 &  $<1.8^{+0.8}_{-0.3}\times 10^ {4}$ &     5,5,5 \\      
Andromeda X &  $01^ {h}06^ {m}34^ {s}~ $+$44^ {\circ}48^ {\prime}16^ {\prime \prime}$ &                   -164 &                          670$^{+25}_{-40}$ &  HI4PI &         Unres. &                   11 &  $<3.4^{+0.3}_{-0.4}\times 10^ {4}$ &     5,5,5 \\            
NGC 147 &  $00^ {h}33^ {m}12^ {s}~ $+$48^ {\circ}30^ {\prime}32^ {\prime \prime}$ &                   -193 &                          676$^{\pm28}$ &  HI4PI &         Unres. &                   21 &  $<6.7^{\pm0.6}\times 10^ {4}$ &     5,5,5 \\      
Andromeda XXX &  $00^ {h}36^ {m}35^ {s}~ $+$49^ {\circ}38^ {\prime}48^ {\prime \prime}$ &                   -140 &                          682$^{+31}_{-82}$ &  HI4PI &         Unres. &                   12 &  $<4.0^{+0.4}_{-1.0}\times 10^ {4}$ &     5,5,5 \\    
Andromeda XVII &  $00^ {h}37^ {m}07^ {s}~ $+$44^ {\circ}19^ {\prime}20^ {\prime \prime}$ &                   -252 &                          728$^{+37}_{-27}$ &  HI4PI &         Unres. &                   11 &  $<4.2^{+0.4}_{-0.3}\times 10^ {4}$ &     5,5,5 \\     
Andromeda XXIX &  $23^ {h}58^ {m}56^ {s}~ $+$30^ {\circ}45^ {\prime}20^ {\prime \prime}$ &                   -194 &                          731$^{\pm74}$ &  GALFA1 &         Unres. &                   36 &  $<2.5^{\pm0.5}\times 10^ {4}$ &     5,5,5 \\       
Andromeda XI &  $00^ {h}46^ {m}20^ {s}~ $+$33^ {\circ}48^ {\prime}05^ {\prime \prime}$ &                   -420 &                          735$^{\pm17}$ &  HI4PI &         Unres. &                   12 &  $<4.6^{\pm0.2}\times 10^ {4}$ &     5,5,5 \\       
Andromeda XX &  $00^ {h}07^ {m}31^ {s}~ $+$35^ {\circ}07^ {\prime}56^ {\prime \prime}$ &                   -456 &                          741$^{+41}_{-55}$ &  GALFA2 &         Unres. &                   76 &  $<5.4^{+0.6}_{-0.8}\times 10^ {4}$ &     5,5,5 \\        
Andromeda I &  $00^ {h}45^ {m}40^ {s}~ $+$38^ {\circ}02^ {\prime}28^ {\prime \prime}$ &                   -376 &                          745$^{\pm24}$ &  HI4PI &         Unres. &                   19 &  $<7.5^{\pm0.5}\times 10^ {4}$ &     5,5,5 \\      
Andromeda III &  $00^ {h}35^ {m}34^ {s}~ $+$36^ {\circ}29^ {\prime}52^ {\prime \prime}$ &                   -344 &                          748$^{\pm24}$ &  GALFA2 &         Unres. &                   86 &  $<6.2^{\pm0.4}\times 10^ {4}$ &     5,5,5 \\            
IC 1613 &  $01^ {h}04^ {m}48^ {s}~ $+$02^ {\circ}07^ {\prime}04^ {\prime \prime}$ &                   -232 &                          755$^{\pm42}$ &      - &              - &                    - &     $6.5^{\pm0.7}\times 10^ {7}$ &     5,5,5 \\             
Cetus  &  $00^ {h}26^ {m}11^ {s}~ $-$11^ {\circ}02^ {\prime}40^ {\prime \prime}$ &                    -84 &                          755$^{\pm24}$ &  HI4PI &         Unres. &                   24 &  $<9.5^{\pm0.6}\times 10^ {4}$ &     5,5,5 \\          
Lacerta I$^{24}$ &  $22^ {h}58^ {m}16^ {s}~ $+$41^ {\circ}17^ {\prime}28^ {\prime \prime}$ &                   -198 &                          759$^{\pm42}$ &  HI4PI &         Unres. &                   24 &  $<9.6^{\pm1.1}\times 10^ {4}$ &    5,23,5 \\      
Andromeda VII &  $23^ {h}26^ {m}32^ {s}~ $+$50^ {\circ}40^ {\prime}33^ {\prime \prime}$ &                   -307 &                          762$^{\pm35}$ &  HI4PI &         Unres. &                   16 &  $<6.6^{\pm0.6}\times 10^ {4}$ &     5,5,5 \\     
Andromeda XXVI &  $00^ {h}23^ {m}46^ {s}~ $+$47^ {\circ}54^ {\prime}58^ {\prime \prime}$ &                   -262 &                          762$^{\pm42}$ &  HI4PI &         Unres. &                   12 &  $<5.0^{\pm0.6}\times 10^ {4}$ &     5,5,5 \\       
Andromeda IX &  $00^ {h}52^ {m}53^ {s}~ $+$43^ {\circ}11^ {\prime}45^ {\prime \prime}$ &                   -209 &                          766$^{\pm25}$ &  HI4PI &         Unres. &                   13 &  $<5.5^{\pm0.4}\times 10^ {4}$ &     5,5,5 \\               
LGS3 &  $01^ {h}03^ {m}55^ {s}~ $+$21^ {\circ}53^ {\prime}06^ {\prime \prime}$ &                   -287 &                          769$^{\pm25}$ &      - &              - &                    - &     $3.8^{\pm0.2}\times 10^ {5}$ &     5,5,5 \\    
Andromeda XXIII &  $01^ {h}29^ {m}22^ {s}~ $+$38^ {\circ}43^ {\prime}08^ {\prime \prime}$ &                   -238 &                          769$^{\pm46}$ &  HI4PI &         Unres. &                   20 &  $<8.2^{\pm1}\times 10^ {4}$ &     5,5,5 \\          
Perseus I$^{25}$ &  $03^ {h}01^ {m}24^ {s}~ $+$40^ {\circ}59^ {\prime}18^ {\prime \prime}$ &                   -326 &                          773$^{\pm64}$ &  HI4PI &         Unres. &                   19 &  $<7.9^{\pm1.3}\times 10^ {4}$ &     5,23,5 \\        
Andromeda V &  $01^ {h}10^ {m}17^ {s}~ $+$47^ {\circ}37^ {\prime}41^ {\prime \prime}$ &                   -403 &                          773$^{\pm29}$ &  HI4PI &         Unres. &                   19 &  $<8.0^{\pm0.6}\times 10^ {4}$ &     5,5,5 \\     
Cassiopeia III$^{26}$ &  $00^ {h}35^ {m}59^ {s}~ $+$51^ {\circ}33^ {\prime}35^ {\prime \prime}$ &                   -372 &                          776$^{\pm50}$ &  HI4PI &         Unres. &                   18 &  $<7.6^{\pm1.0}\times 10^ {4}$ &    5,23,5 \\       
Andromeda VI &  $23^ {h}51^ {m}46^ {s}~ $+$24^ {\circ}34^ {\prime}57^ {\prime \prime}$ &                   -340 &                          783$^{\pm25}$ &  GALFA1 &         Unres. &                   30 &  $<2.4^{\pm0.2}\times 10^ {4}$ &     5,5,5 \\      
Andromeda XIV &  $00^ {h}51^ {m}35^ {s}~ $+$29^ {\circ}41^ {\prime}49^ {\prime \prime}$ &                   -481 &                          794$^{+22}_{-205}$ &  GALFA1 &         Unres. &                   29 &  $<2.4^{+0.1}_{-1.2}\times 10^ {4}$ &     5,5,5 \\             
IC 10 &  $00^ {h}20^ {m}17^ {s}~ $+$59^ {\circ}18^ {\prime}14^ {\prime \prime}$ &                   -348 &                          794$^{\pm44}$ &      - &              - &                    - &     $5.0^{\pm0.6}\times 10^ {7}$ &     5,5,5 \\              
Leo A &  $09^ {h}59^ {m}26^ {s}~ $+$30^ {\circ}44^ {\prime}47^ {\prime \prime}$ &                     24 &                          798$^{\pm44}$ &      - &              - &                    - &     $1.1^{\pm0.1}\times 10^ {7}$ &     5,5,5 \\                
M32 &  $00^ {h}42^ {m}42^ {s}~ $+$40^ {\circ}51^ {\prime}55^ {\prime \prime}$ &                   -199 &                          805$^{\pm78}$ &  HI4PI &         Unres. &                   38 &  $<1.8^{\pm0.3}\times 10^ {5}$ &     5,5,5 \\      
Andromeda XXV &  $00^ {h}30^ {m}09^ {s}~ $+$46^ {\circ}51^ {\prime}07^ {\prime \prime}$ &                   -108 &                          813$^{\pm45}$ &  HI4PI &         Unres. &                   62 &  $<2.9^{\pm0.3}\times 10^ {5}$ &     5,5,5 \\      
Andromeda XIX &  $00^ {h}19^ {m}32^ {s}~ $+$35^ {\circ}02^ {\prime}37^ {\prime \prime}$ &                   -112 &                          820$^{+30}_{-162}$ &  GALFA2 &           Res. &                   39 &  $<1.3^{+0.1}_{-0.5}\times 10^ {5}$ &     5,5,5 \\            
NGC 205 &  $00^ {h}40^ {m}22^ {s}~ $+$41^ {\circ}41^ {\prime}07^ {\prime \prime}$ &                   -246 &                          824$^{\pm27}$ &      - &              - &                    - &     $4.0^{\pm0.3}\times 10^ {5}$ &     5,5,5 \\      
Andromeda XXI &  $23^ {h}54^ {m}48^ {s}~ $+$42^ {\circ}28^ {\prime}15^ {\prime \prime}$ &                   -363 &                          828$^{+23}_{-27}$ &  HI4PI &         Unres. &                   14 &  $<6.9^{\pm0.4}\times 10^ {4}$ &     5,5,5 \\    
Andromeda XXVII &  $00^ {h}37^ {m}27^ {s}~ $+$45^ {\circ}23^ {\prime}13^ {\prime \prime}$ &                   -540 &                          828$^{\pm46}$ &  HI4PI &         Unres. &                   19 &  $<9.1^{\pm1}\times 10^ {4}$ &     5,5,5 \\     
Andromeda XIII &  $00^ {h}51^ {m}51^ {s}~ $+$33^ {\circ}00^ {\prime}16^ {\prime \prime}$ &                   -185 &                          840$^{\pm19}$ &  GALFA2 &         Unres. &                   76 &  $<7.0^{\pm0.3}\times 10^ {4}$ &     5,5,5 \\            
Tucana  &  $22^ {h}41^ {m}50^ {s}~ $-$64^ {\circ}25^ {\prime}10^ {\prime \prime}$ &                    194 &                          887$^{\pm49}$ &  HI4PI &         Unres. &                   15 &  $<8.5^{\pm0.9}\times 10^ {4}$ &     5,5,5 \\     
Andromeda XXII &  $01^ {h}27^ {m}40^ {s}~ $+$28^ {\circ}05^ {\prime}25^ {\prime \prime}$ &                   -130 &                          920$^{+30}_{-153}$ &  GALFA1 &         Unres. &                   36 &  $<3.9^{+0.3}_{-1.3}\times 10^ {4}$ &     5,5,5 \\       
Pegasus dIrr &  $23^ {h}28^ {m}36^ {s}~ $+$14^ {\circ}44^ {\prime}35^ {\prime \prime}$ &                   -180 &                          920$^{\pm30}$ &      - &              - &                    - &     $5.9^{\pm0.4}\times 10^ {6}$ &     5,5,5 \\      
Andromeda XII &  $00^ {h}47^ {m}27^ {s}~ $+$34^ {\circ}22^ {\prime}29^ {\prime \prime}$ &                   -558 &                          929$^{+39}_{-145}$ &  GALFA2 &         Unres. &                   71 &  $<7.9^{+0.7}_{-2.5}\times 10^ {4}$ &     5,5,5 \\                
WLM &  $00^ {h}01^ {m}58^ {s}~ $-$15^ {\circ}27^ {\prime}39^ {\prime \prime}$ &                   -130 &                          933$^{\pm34}$ &      - &              - &                    - &     $6.1^{\pm0.5}\times 10^ {7}$ &     5,5,5 \\   
Sagittarius dIrr &  $19^ {h}29^ {m}59^ {s}~ $-$17^ {\circ}40^ {\prime}51^ {\prime \prime}$ &                    -79 &                         1067$^{\pm88}$ &      - &              - &                    - &     $8.8^{\pm1.5}\times 10^ {6}$ &     5,5,5 \\          
Aquarius  &  $20^ {h}46^ {m}52^ {s}~ $-$12^ {\circ}50^ {\prime}53^ {\prime \prime}$ &                   -138 &                         1072$^{\pm40}$ &      - &              - &                    - &     $4.1^{\pm0.3}\times 10^ {6}$ &     5,5,5 \\    
Andromeda XVIII &  $00^ {h}02^ {m}14^ {s}~ $+$45^ {\circ}05^ {\prime}20^ {\prime \prime}$ &                   -332 &                         1213$^{+39}_{-45}$ &  HI4PI &         Unres. &                   21 &  $<2.2^{+0.1}_{-0.2}\times 10^ {5}$ &     5,5,5 \\           
Antlia B &  $09^ {h}48^ {m}56^ {s}~ $-$25^ {\circ}59^ {\prime}24^ {\prime \prime}$ &                    376 &                         1294$^{\pm95}$ &      - &              - &                    - &     $2.8^{\pm0.4}\times 10^ {5}$ &     5,5,5 \\           
NGC 3109 &  $10^ {h}03^ {m}07^ {s}~ $-$26^ {\circ}09^ {\prime}35^ {\prime \prime}$ &                    403 &                         1300$^{\pm48}$ &      - &              - &                    - &     $4.5^{\pm0.3}\times 10^ {8}$ &     5,5,5 \\            
Antlia  &  $10^ {h}04^ {m}04^ {s}~ $-$27^ {\circ}19^ {\prime}52^ {\prime \prime}$ &                    362 &                         1349$^{\pm62}$ &      - &              - &                    - &     $7.3^{\pm0.7}\times 10^ {5}$ &     5,5,5\\           
UGC 4879 &  $09^ {h}16^ {m}02^ {s}~ $+$52^ {\circ}50^ {\prime}24^ {\prime \prime}$ &                    -29 &                         1361$^{\pm25}$ &      - &              - &                    - &     $9.5^{\pm0.4}\times 10^ {5}$ &     5,5,5 \\          
Sextans B &  $10^ {h}00^ {m}00^ {s}~ $+$05^ {\circ}19^ {\prime}56^ {\prime \prime}$ &                    304 &                         1426$^{\pm20}$ &      - &              - &                    - &     $5.1^{\pm0.1}\times 10^ {7}$ &     5,5,5 \\          
Sextans A &  $10^ {h}11^ {m}01^ {s}~ $-$04^ {\circ}41^ {\prime}34^ {\prime \prime}$ &                    324 &                         1432$^{\pm53}$ &      - &              - &                    - &     $7.7^{\pm0.6}\times 10^ {7}$ &     5,5,5 \\              
Leo P &  $10^ {h}21^ {m}45^ {s}~ $+$18^ {\circ}05^ {\prime}17^ {\prime \prime}$ &                    264 &                         1622$^{\pm149}$ &      - &              - &                    - &     $9.4^{\pm1.7}\times 10^ {5}$ &     5,5,5 \\            
HIZSS3B &  $07^ {h}00^ {m}29^ {s}~ $-$04^ {\circ}12^ {\prime}30^ {\prime \prime}$ &                    323 &                         1675$^{\pm108}$ &      - &              - &                    - &     $2.6^{\pm0.3}\times 10^ {6}$ &     5,5,5 \\          
HIZSS3(A) &  $07^ {h}00^ {m}29^ {s}~ $-$04^ {\circ}12^ {\prime}30^ {\prime \prime}$ &                    288 &                         1675$^{\pm108}$ &      - &              - &                    - &     $1.4^{\pm0.2}\times 10^ {7}$ &     5,5,5 \\              
KKR25 &  $16^ {h}13^ {m}48^ {s}~ $+$54^ {\circ}22^ {\prime}16^ {\prime \prime}$ &                    -65 &                         1923$^{\pm62}$ &  HI4PI &         Unres. &                   23 &  $<6.1^{\pm0.4}\times 10^ {5}$ &     5,5,5 \\        
ESO410-G005 &  $00^ {h}15^ {m}32^ {s}~ $-$32^ {\circ}10^ {\prime}48^ {\prime \prime}$ &                    - &                         1923$^{\pm35}$ &      - &              - &                    - &     $7.3^{\pm0.3}\times 10^ {5}$ &     5,27,5 \\            
IC 5152 &  $22^ {h}02^ {m}42^ {s}~ $-$51^ {\circ}17^ {\prime}47^ {\prime \prime}$ &                    122 &                         1950$^{\pm45}$ &      - &              - &                    - &     $8.7^{\pm0.4}\times 10^{7}$ &     5,5,5 \\
\end{longtable*}
\end{ThreePartTable}
\end{center} 
\clearpage

\begin{center}
\begin{ThreePartTable}
\begin{TableNotes}
\label{tab:other}
\item Notes -- 
(1) The parameters use the distances and coordinates from Table 1 where relevant.
(2)  When $\sigma_{gas}$ is noted it is derived directly from the FWHM of the HI line.
(3) This is the distance from the Local Group surface defined in \S3.1, while D$_{\rm MW}$ and D$_{\rm M31}$ are from the center of the MW and M31.  The value in superscript is the variation on the distance from this surface when the Milky Way mass is higher and the M31 mass is lower.  The subscript value is when the Milky Way mass is lower and the M31 mass is higher.  The total Local Group mass remains the same in all cases.
(4) This column notes the references for the luminosity, half-light radius, and velocity dispersion, respectively.
(5) \cite{mcconnachie12}.
(6) \cite{simon19}.
(7) $\sigma$ is from \cite{martinn16}.
(8) Properties are from \cite{drlica16}.
(9) $\sigma$ is from \cite{sand18}.  No r$_{\rm h}$ is given since this galaxy is highly disrupted.
(10) $\sigma$ is from \cite{simon20}.
(11)  r$_{\rm h}$ is from \cite{bothun88}. 
(12) The LMC has a very large stellar rotation component compared to the stellar dispersion component.  Using the stellar dispersion value (20.2 \kms), would lead to a severe underestimate of the total mass, so we use the gas dispersion calculated from the FWHM of the HI line.
(13) Properties are from \cite{homma16}.
(14) Properties are from \cite{torrealba19}.
(15) Properties are from \cite{homma18}.
(16) $\sigma$ is from \cite{kacharov17}.
(17) $\sigma$ is from \cite{martin14}.
(18) $\sigma$ is from \cite{namumba19}.
(19) r$_{\rm h}$ is from \cite{martin16}. $\sigma$ is from \citep{collins20}.
(20) $\sigma$ is from \cite{taibi20}.
(21) $\sigma$ is from \cite{sand15}.
(22) $\sigma$ is from \cite{barnes01}.
(23) $\sigma$ is from \cite{namumba18}.
(24) $\sigma$ is from \cite{bernstein14}. r$_{\rm h}$ in this case is the semi-major axis noted by \cite{mcquinn15}.
(25) $\sigma$ is from \cite{begum05}.
(26) $\sigma$ is from \cite{bouchard05} and is consistent with the detection at the lower velocity by \cite{koribalski18}. 
(27) $\sigma$ is from \cite{koribalski18}.

\end{TableNotes}
\begin{longtable*}{llrccccc}
\caption{Dwarf Galaxies within 2 Mpc:  Other Parameters$^{1}$}\\
\hline
Galaxy & L$_{\rm V}$ (L$_\odot$) & r$_{\rm h}$ (pc) & $\sigma_*$ (\kms)$^2$ & D$_{\rm MW}$ (kpc) & D$_{\rm M31}$ (kpc) & D$_{\rm LG}$ (kpc)$^3$ & Refs.$^4$ \\
\hline
\endfirsthead
\hline
Galaxy & L$_{\rm V}$ (L$_\odot$) & r$_{\rm h}$ (pc) & $\sigma_*$ (\kms)$^2$ & D$_{\rm MW}$ (kpc) & D$_{\rm M31}$ (kpc) & D$_{\rm LG}$ (kpc)$^3$ & Refs.$^4$ \\
\hline
\endhead
\endfoot
\hline
\insertTableNotes
\endlastfoot
Draco II & $1.8\times10^{2}$ & 19 & 2.9 & 24 & 771 & $-210^{+15}_{-15}$ & 6,6,7 \\
Segue 1 & $2.8\times10^{2}$ & 24 & 3.7 & 28 & 788 & $-197^{+18}_{-17}$ & 6,6,6\\
DESJ0225+0304 & $2.4\times10^{2}$ & 19 & - & 29 & 762 & $-214^{+14}_{-15}$ & 6,6,-\\
Tucana III & $3.4\times10^{2}$ & 37 & $<$1.2  & 23 & 784 & $-203^{+18}_{-17}$ & 6,6,6\\
Sagittarius dSph & $2.1\times10^{7}$ & 2662 & 11.4 & 19 & 787 & $-206^{+19}_{-17}$ & 6,6,6\\
Hydrus I & $6.5\times10^{3}$ & 53 & 2.7 & 26 & 794 & $-198^{+20}_{-17}$ & 6,6,6\\
Carina III & $7.8\times10^{2}$ & 30 & 5.6 & 29 & 797 & $-195^{+20}_{-18}$ & 6,6,6 \\
Triangulum II & $3.7\times10^{2}$ & 16 & $<$3.4 & 35 & 752 & $-222^{+12}_{-14}$ & 6,6,6 \\
Cetus II & $8.6\times10^{1}$ & 17 & - & 32 & 764 & $-208^{+13}_{-14}$ & 6,6,-\\
Reticulum 2 & $3.0\times10^{3}$ & 51 & 3.3 & 33 & 786 & $-193^{+17}_{-16}$ & 6,6,6\\
Ursa Major II & $5.1\times10^{3}$ & 139 & 5.6 & 41 & 766 & $-198^{+13}_{-14}$ & 6,6,6\\
Carina II & $5.4\times10^{3}$ & 92 & 3.4 & 37 & 803 & $-187^{+20}_{-18}$ & 6,6,6\\
Segue II & $5.3\times10^{2}$ & 40 & 3.4 & 43 & 747 & $-217^{+11}_{-13}$ & 6,6,6\\
Coma Berenices & $4.4\times10^{3}$ & 69 & 4.6 & 43 & 797 & $-181^{+18}_{-17}$ & 6,6,6\\
Bootes II & $1.3\times10^{3}$ & 39 & 10.5 & 40 & 803 & $-184^{+20}_{-18}$ & 6,6,6\\
Pictor II & $1.6\times10^{3}$ & 47 & - & 46 & 806 & $-179^{+20}_{-18}$ & 8,8,- \\
Willman 1 & $1.2\times10^{3}$ & 33 & 4.0 & 50 & 776 & $-183^{+13}_{-15}$ & 6,6,6\\
Bootes III & $1.7\times10^{4}$ & - & 10.7 & 46 & 796 & $-179^{+17}_{-17}$ & 5,-,9\\
Tucana IV & $2.1\times10^{3}$ & 127 & 4.3 & 45 & 791 & $-180^{+16}_{-16}$ & 6,6,10\\
LMC & $1.5\times10^{9}$ & 2675 & $\sigma_{gas}=$34   & 50 & 807 & $-174^{+19}_{-18}$ & 5,11,12\\
Grus II & $3.1\times10^{3}$ & 93 & $<$1.9 & 48 & 785 & $-180^{+15}_{-16}$ & 6,6,10\\
Tucana V & $3.7\times10^{2}$ & 16 & - & 52 & 795 & $-173^{+16}_{-16}$ & 6,6,-\\
Tucana II & $3.1\times10^{3}$ & 121 & 8.6 & 54 & 794 & $-172^{+16}_{-16}$ & 6,6,6\\
SMC & $4.6\times10^{8}$ & 1105 & 27.6 & 61 & 807 & $-163^{+18}_{-17}$ & 5,11,5\\
Bootes & $2.2\times10^{4}$ & 191 & 4.6 & 64 & 815 & $-161^{+20}_{-18}$ & 6,6,6\\
Sagittarius II & $1.0\times10^{4}$ & 33 & - & 63 & 785 & $-166^{+13}_{-15}$ & 6,6,-\\
Ursa Minor & $3.5\times10^{5}$ & 405 & 9.5 & 78 & 754 & $-171^{+8}_{-12}$ & 6,6,6\\
Horologium II & $3.6\times10^{2}$ & 44 & - & 79 & 793 & $-150^{+13}_{-15}$ & 6,6,-\\
Draco & $3.0\times10^{5}$ & 231 & 9.1 & 82 & 748 & $-171^{+7}_{-12}$ & 6,6,6\\
Phoenix 2 & $1.0\times10^{3}$ & 37 & - & 81 & 793 & $-147^{+13}_{-15}$ & 6,6,-\\
Sculptor & $1.8\times10^{6}$ & 279 & 9.2 & 86 & 761 & $-158^{+8}_{-12}$ & 6,6,6\\
Horologium 1 & $2.7\times10^{3}$ & 40 & 4.9 & 87 & 798 & $-141^{+13}_{-15}$ & 6,6,6\\
Virgo I & $1.8\times10^{2}$ & 38 & - & 87 & 845 & $-141^{+24}_{-21}$ & 13,13,-\\
Eridanus 3 & $5.4\times10^{2}$ & 14 & - & 87 & 793 & $-142^{+12}_{-15}$ & 5,5,-\\
Reticulum III & $1.8\times10^{3}$ & 64 & - & 92 & 813 & $-133^{+15}_{-16}$ & 6,6,-\\
Sextans & $3.2\times10^{5}$ & 456 & 7.9 & 98 & 841 & $-127^{+21}_{-19}$ & 6,6,6\\
Ursa Major & $9.6\times10^{3}$ & 295 & 7.0 & 102 & 773 & $-137^{+8}_{-13}$ & 6,6,6\\
Indus I & $2.1\times10^{3}$ & 37 & - & 94 & 808 & $-132^{+14}_{-16}$ & 5,5,-\\
Carina & $5.2\times10^{5}$ & 311 & 6.6 & 108 & 838 & $-117^{+19}_{-18}$ & 6,6,6\\
Aquarius II & $4.7\times10^{3}$ & 160 & 5.4 & 105 & 729 & $-164^{+3}_{-10}$ & 6,6,6\\
Crater II & $1.6\times10^{5}$ & 1066 & 2.7 & 116 & 886 & $-126^{+35}_{-26}$ & 6,6,6\\
Grus I & $2.1\times10^{3}$ & 28 & 2.9 & 116 & 797 & $-115^{+10}_{-14}$ & 6,6,6\\
Pictoris 1 & $2.5\times10^{3}$ & 32 & - & 128 & 822 & $-99^{+13}_{-16}$ & 6,6,-\\
Antlia 2 & $3.5\times10^{5}$ & 2899 & 5.7 & 133 & 889 & $-101^{+30}_{-25}$ &  14,14,14\\
Hercules & $1.8\times10^{4}$ & 216 & 5.1 & 126 & 822 & $-100^{+13}_{-16}$ & 6,6,6\\
Fornax & $1.9\times10^{7}$ & 792 & 11.7 & 141 & 768 & $-105^{+5}_{-11}$ & 6,6,6\\
Hydra II & $7.5\times10^{3}$ & 67 & $<$3.6 & 148 & 928 & $-118^{+52}_{-36}$ & 6,6,6\\
Leo IV & $8.5\times10^{3}$ & 114 & 3.3 & 155 & 895 & $-75^{+26}_{-23}$ & 6,6,6\\
Canes Venatici II & $1.0\times10^{4}$ & 71 & 4.6 & 161 & 833 & $-66^{+12}_{-15}$ & 6,6,6\\
Leo V & $4.4\times10^{3}$ & 49 & 2.3 & 170 & 904 & $-60^{+26}_{-23}$ & 6,6,6\\
Pisces II & $4.2\times10^{3}$ & 60 & 5.4 & 182 & 655 & $-154^{-4}_{-6}$ & 6,6,6\\
Columba I & $4.1\times10^{3}$ & 117 & - & 188 & 819 & $-45^{+7}_{-13}$ & 6,6,-\\
Pegasus III & $3.7\times10^{3}$ & 78 & 5.4 & 203 & 657 & $-128^{-4}_{-7}$ & 6,6,6\\
Canes Venatici I & $2.7\times10^{5}$ & 437 & 7.6 & 211 & 856 & $-16^{+10}_{-15}$ & 6,6,6\\
Indus II & $4.5\times10^{3}$ & 181 & - & 208 & 853 & $-20^{+10}_{-15}$ & 6,6,-\\
Leo II & $6.7\times10^{5}$ & 171 & 7.4 & 236 & 897 & $11^{+14}_{-17}$ & 6,6,6\\
Cetus III & $8.2\times10^{2}$ & 90 & - & 255 & 644 & $-89^{-2}_{-10}$ & 15,15,-\\
Leo I & $4.4\times10^{6}$ & 270 & 9.2 & 258 & 918 & $34^{+15}_{-18}$ & 6,6,6\\
Eridanus II & $5.9\times10^{4}$ & 246 & 6.9 & 368 & 884 & $128^{+1}_{-10}$ & 6,6,6\\
Leo T & $1.4\times10^{5}$ & 118 & 7.5 & 414 & 982 & $187^{+8}_{-15}$ & 6,6,6\\
Phoenix & $7.7\times10^{5}$ & 454 & 9.3 & 415 & 864 & $160^{-7}_{-4}$ & 5,5,16\\
NGC 6822 & $1.0\times10^{8}$ & 354 & 23.2 & 452 & 894 & $200^{-7}_{-4}$ & 5,5,5\\
Andromeda XVI & $3.4\times10^{5}$ & 123 & 3.8 & 480 & 319 & $-231^{+12}_{-29}$ & 5,5,5\\
Andromeda XXIV & $9.3\times10^{4}$ & 366 & - & 605 & 204 & $-219^{+17}_{-34}$ & 5,5,-\\
NGC 185 & $6.8\times10^{7}$ & 457 & 24.0 & 621 & 184 & $-228^{+18}_{-35}$ & 5,5,5\\
Andromeda XV & $4.8\times10^{5}$ & 220 & 4.0 & 630 & 175 & $-229^{+18}_{-35}$ & 5,5,5\\
Andromeda II & $9.1\times10^{6}$ & 1175 & 7.8 & 656 & 181 & $-185^{+19}_{-36}$ & 5,5,5\\
Andromeda XXVIII & $2.1\times10^{5}$ & 213 & 6.6 & 661 & 365 & $0^{+16}_{-34}$ & 5,5,5\\
Andromeda X & $8.8\times10^{4}$ & 253 & 6.4 & 674 & 130 & $-233^{+20}_{-37}$ & 5,5,5\\
NGC 147 & $6.2\times10^{7}$ & 623 & 16.0 & 680 & 139 & $-212^{+20}_{-37}$ & 5,5,5\\
Andromeda XXX & $1.3\times10^{5}$ & 268 & 11.8 & 686 & 144 & $-199^{+20}_{-37}$ & 5,5,5\\
Andromeda XVII & $2.2\times10^{5}$ & 263 & 2.9 & 732 & 66 & $-251^{+23}_{-40}$ & 5,5,5\\
Andromeda XXIX & $1.8\times10^{5}$ & 362 & 5.7 & 734 & 187 & $-123^{+21}_{-38}$ & 5,5,5\\
Andromeda XI & $4.6\times10^{4}$ & 152 & $<$4.6 & 738 & 108 & $-196^{+23}_{-39}$ & 5,5,5\\
Andromeda XX & $2.5\times10^{4}$ & 114 & 7.1 & 744 & 128 & $-173^{+23}_{-39}$ & 5,5,5\\
Andromeda I & $4.7\times10^{6}$ & 672 & 10.2 & 749 & 55 & $-247^{+24}_{-40}$ & 5,5,5\\
Andromeda III & $1.0\times10^{6}$ & 479 & 9.3 & 752 & 73 & $-224^{+24}_{-40}$ & 5,5,5\\
IC 1613 & $1.0\times10^{8}$ & 1496 & 10.8 & 758 & 518 & $172^{+15}_{-33}$ & 5,5,5 \\
Cetus & $2.8\times10^{6}$ & 703 & 8.3 & 756 & 678 & $288^{+11}_{-29}$ & 5,5,5\\
Lacerta I & $4.1\times10^{6}$ & 927 & 10.3 & 760 & 262 & $-42^{+21}_{-38}$ & 5,5,17\\
Andromeda XXVI & $6.0\times10^{4}$ & 222 & 8.6 & 766 & 102 & $-189^{+24}_{-40}$ & 5,5,5\\
Andromeda VII & $1.6\times10^{7}$ & 776 & 13.0 & 765 & 217 & $-81^{+22}_{-39}$ & 5,5,5\\
Andromeda IX & $1.5\times10^{5}$ & 557 & 10.9 & 770 & 39 & $-249^{+25}_{-41}$ & 5,5,5\\
Andromeda XXIII & $1.1\times10^{6}$ & 1029 & 7.1 & 774 & 126 & $-163^{+24}_{-40}$ & 5,5,5\\
LGS3 & $9.6\times10^{5}$ & 470 & 7.9 & 773 & 268 & $-33^{+21}_{-39}$ & 5,5,5\\
Perseus I & $1.2\times10^{6}$ & 382 & 4.2 & 779 & 348 & $41^{+20}_{-37}$ & 5,5,17\\
Andromeda V & $5.6\times10^{5}$ & 315 & 11.5 & 777 & 109 & $-178^{+25}_{-41}$ & 5,5,5\\
Cassiopeia III & $6.8\times10^{6}$ & 1468 & 8.4 & 780 & 140 & $-147^{+24}_{-40}$ & 5,5,17\\
Andromeda VI & $3.3\times10^{6}$ & 524 & 12.4 & 785 & 268 & $-28^{+22}_{-39}$ & 5,5,5\\
IC 10 & $8.6\times10^{7}$ & 612 & $\sigma_{gas}=$25.9 & 798 & 252 & $-40^{+23}_{-40}$ & 5,5,18\\
Andromeda XIV & $2.4\times10^{5}$ & 393 & 5.3 & 798 & 161 & $-124^{+25}_{-41}$ & 5,5,5\\
Leo A & $6.0\times10^{6}$ & 499 & 6.7 & 803 & 1197 & $564^{-9}_{-3}$ & 5,5,5\\
M32 & $3.2\times10^{8}$ & 110 & 92.0 & 809 & 27 & $-271^{+28}_{-43}$ & 5,5,5\\
Andromeda XXV & $6.8\times10^{5}$ & 709 & 3.0 & 817 & 90 & $-192^{+27}_{-42}$ & 5,5,5\\
Andromeda XIX & $3.3\times10^{5}$ & 3065 & 7.8 & 824 & 115 & $-167^{+27}_{-42}$ & 5,19,19 \\
NGC 205 & $3.3\times10^{8}$ & 590 & 35.0 & 828 & 46 & $-263^{+29}_{-43}$ & 5,5,5\\
Andromeda XXVII & $1.2\times10^{5}$ & 433 & 14.8 & 832 & 77 & $-212^{+28}_{-43}$ & 5,5,5\\
Andromeda XXI & $7.0\times10^{5}$ & 843 & 4.5 & 831 & 135 & $-147^{+27}_{-42}$ & 5,5,5\\
Andromeda XIII & $3.5\times10^{4}$ & 191 & 5.8 & 843 & 134 & $-150^{+28}_{-43}$ & 5,5,5\\
Tucana & $5.6\times10^{5}$ & 284 & 6.2 & 883 & 1352 & $657^{+6}_{-15}$ & 5,5,20\\
Andromeda XXII & $4.6\times10^{4}$ & 252 & 2.8 & 925 & 276 & $-11^{+29}_{-44}$ & 5,5,5\\
Pegasus dIrr & $6.6\times10^{6}$ & 562 & 12.3 & 921 & 474 & $187^{+22}_{-40}$ & 5,5,5\\
Andromeda XII & $3.5\times10^{4}$ & 324 & 2.6 & 933 & 182 & $-137^{+34}_{-46}$ & 5,5,5\\
WLM & $4.3\times10^{7}$ & 2340 & 17.5 & 933 & 835 & $476^{+12}_{-31}$ & 5,5,5\\
Sagittarius dIrr & $3.5\times10^{6}$ & 282 & $\sigma_{gas}=$10 & 1059 & 1354 & $806^{-5}_{-9}$ & 5,5,5\\
Aquarius & $1.6\times10^{6}$ & 458 & 7.9 & 1066 & 1170 & $740^{+5}_{-22}$ & 5,5,5\\
Andromeda XVIII & $5.0\times10^{5}$ & 325 & 9.7 & 1217 & 457 & $44^{+56}_{-58}$ & 5,5,5\\
Antlia B & $6.3\times10^{5}$ & 271 & $\sigma_{gas}=$7.2 & 1296 & 1963 & $1047^{+43}_{-45}$ & 5,5,21\\
NGC 3109 & $7.6\times10^{7}$ & 1626 & $\sigma_{gas}=$61 & 1301 & 1984 & $1049^{+46}_{-47}$ & 5,5,22\\
Antlia & $1.3\times10^{6}$ & 471 & $\sigma_{gas}=$12.7 & 1350 & 2036 & $1096^{+47}_{-48}$ & 5,5,22\\
UGC 4879 & $8.3\times10^{6}$ & 162 & 9.6 & 1367 & 1394 & $1017^{+8}_{-26}$ & 5,5,5\\
Sextans B & $5.2\times10^{7}$ & 440 & $\sigma_{gas}=$17 & 1429 & 1940 & $1204^{+18}_{-25}$ & 5,5,23\\
Sextans A & $4.4\times10^{7}$ & 1029 & $\sigma_{gas}=$19 & 1435 & 2024 & $1201^{+30}_{-35}$ & 5,5,23\\
Leo P & $3.9\times10^{5}$ & 566 & $\sigma_{gas}=$8.4 & 1625 & 2048 & $1401^{+8}_{-18}$ & 5,24,24\\
HIZSS3B & - & - & $\sigma_{gas}=$11.9 & 1681 & 1921 & $1429^{-4}_{-11}$ & -,-,25\\
HIZSS3(A) & - & - & $\sigma_{gas}=$23 & 1681 & 1921 & $1429^{-4}_{-11}$ & -,-,25\\
KKR25 & $2.0\times10^{6}$ & 263 & - & 1922 & 1869 & $1541^{+12}_{-31}$ & 5,5,-\\
ESO410-G005 & $3.5\times10^{6}$ & 280 & $\sigma_{gas}=$14.2 & 1922 & 1861 & $1536^{+12}_{-32}$ & 5,5,26\\
IC 5152 & $1.4\times10^{8}$ & 550 & $\sigma_{gas}=$36 & 1945 & 2209 & $1702^{-6}_{-8}$ & 5,5,27\\

\end{longtable*}
\end{ThreePartTable}
\end{center}

\clearpage

\begin{center}
\begin{ThreePartTable}
\begin{TableNotes}
\label{tab:orbit}
\item Note -- 
(1) This is the median halo density required to strip the dwarf at this position.  The errors should be multiplied by the same factor of 10 noted for the median density.
(2) This galaxy is not included in Figures 8 and 9 because of its very small pericenter.
(3) Phoenix is the only galaxy in this list with gas.
\end{TableNotes}
\begin{longtable*}{lllllll}
\caption{Median Orbital Parameters and Halo Densities}\\
\hline
Galaxy & r$_{peri}$ (kpc) & $v_{peri}$ (\kms) &        $n_{halo, peri}$ (cm$^ {-3}$)$^1$ &     r$_{apo}$ (kpc) & $v_{apo}$ (\kms) &         $n_{halo, apo}$ (cm$^ {-3}$)$^1$ \\ 
\hline
\endfirsthead
\hline
Galaxy & r$_{peri}$ (kpc) & $v_{peri}$ (\kms) &        $n_{halo, peri}$ (cm$^ {-3}$)$^1$ &     r$_{apo}$ (kpc) & $v_{apo}$ (\kms) &         $n_{halo, apo}$ (cm$^ {-3}$)$^1$ \\ 
\hline
\endhead
\endfoot
\hline
\insertTableNotes
\endlastfoot
\midrule   
Tucana III$^2$ & 3$^{+0.3}_{-0.3}$ & 536$^{\pm19}$ & 3.5$^{+11.1}_{-3.2}\times10^{-7}$ & 37$^{+4}_{-3}$ & 34$^{\pm5}$ & 8.8$^{+28.7}_{-8.0}\times10^{-5}$ \\
Sagittarius dSph & 15$^{+2}_{-2}$ & 354$^{\pm9}$ & 3.8$^{+0.5}_{-0.5}\times10^{-4}$ & 50$^{+19}_{-11}$ & 106$^{\pm20}$ & 4.2$^{+2.8}_{-1.3}\times10^{-3}$ \\
Triangulum II & 17$^{+3}_{-2}$ & 426$^{\pm29}$ & 5.9$^{+11.8}_{-5.3}\times10^{-6}$ & 146$^{+81}_{-36}$ & 50$^{\pm14}$ & 4.3$^{+9.6}_{-3.9}\times10^{-4}$ \\
Segue 1 & 18$^{+5}_{-6}$ & 321$^{\pm25}$ & 4.8$^{+4.1}_{-3.0}\times10^{-5}$ & 41$^{+17}_{-8}$ & 128$^{\pm18}$ & 3.1$^{+3.2}_{-2.0}\times10^{-4}$ \\
Willman 1 & 19$^{+16}_{-8}$ & 306$^{\pm64}$ & 6.3$^{+4.8}_{-2.8}\times10^{-5}$ & 43$^{+9}_{-7}$ & 139$^{\pm35}$ & 3.1$^{+3.0}_{-1.4}\times10^{-4}$ \\
Draco II & 19$^{+1}_{-1}$ & 378$^{\pm16}$ & 2.7$^{+4.2}_{-2.2}\times10^{-5}$ & 97$^{+53}_{-27}$ & 77$^{\pm23}$ & 6.5$^{+15.4}_{-5.5}\times10^{-4}$ \\
Hydrus I & 25$^{+1}_{-1}$ & 372$^{\pm10}$ & 1.9$^{+0.8}_{-0.7}\times10^{-5}$ & 128$^{+85}_{-36}$ & 73$^{\pm25}$ & 5.3$^{+8.8}_{-2.8}\times10^{-4}$ \\
Reticulum 2 & 25$^{+3}_{-4}$ & 300$^{\pm13}$ & 4.4$^{+2.1}_{-1.7}\times10^{-5}$ & 53$^{+14}_{-10}$ & 139$^{\pm19}$ & 2.1$^{+1.3}_{-0.9}\times10^{-4}$ \\
Carina II & 27$^{+1}_{-2}$ & 397$^{\pm9}$ & 2.8$^{+2.3}_{-1.5}\times10^{-5}$ & 219$^{+176}_{-71}$ & 54$^{\pm19}$ & 1.5$^{+2.6}_{-0.9}\times10^{-3}$ \\
Crater II & 27$^{+14}_{-10}$ & 362$^{\pm60}$ & 2.0$^{+1.0}_{-0.7}\times10^{-5}$ & 128$^{+9}_{-7}$ & 77$^{\pm20}$ & 4.5$^{+4.2}_{-1.8}\times10^{-4}$ \\
Carina III & 28$^{+1}_{-1}$ & 392$^{\pm28}$ & 8.7$^{+15.9}_{-7.2}\times10^{-5}$ & 228$^{+306}_{-106}$ & 65$^{\pm37}$ & 3.0$^{+9.6}_{-2.6}\times10^{-3}$ \\
Tucana II & 33$^{+25}_{-9}$ & 358$^{\pm28}$ & 2.2$^{+2.9}_{-1.7}\times10^{-4}$ & 160$^{+142}_{-55}$ & 73$^{\pm22}$ & 5.1$^{+9.6}_{-3.9}\times10^{-3}$ \\
Draco & 33$^{+8}_{-6}$ & 313$^{\pm34}$ & 3.1$^{+1.2}_{-0.9}\times10^{-4}$ & 96$^{+11}_{-9}$ & 109$^{\pm9}$ & 2.6$^{+0.9}_{-0.7}\times10^{-3}$ \\
Segue II & 33$^{+5}_{-7}$ & 275$^{\pm18}$ & 6.2$^{+11.6}_{-4.9}\times10^{-5}$ & 57$^{+22}_{-10}$ & 153$^{\pm22}$ & 2.2$^{+4.2}_{-1.7}\times10^{-4}$ \\
Ursa Minor & 34$^{+7}_{-6}$ & 302$^{\pm34}$ & 3.7$^{+1.4}_{-1.1}\times10^{-4}$ & 88$^{+6}_{-5}$ & 117$^{\pm10}$ & 2.4$^{+0.8}_{-0.7}\times10^{-3}$ \\
Hercules & 36$^{+36}_{-22}$ & 361$^{\pm82}$ & 6.9$^{+4.8}_{-3.1}\times10^{-5}$ & 232$^{+138}_{-40}$ & 57$^{\pm28}$ & 2.9$^{+6.1}_{-1.5}\times10^{-3}$ \\
Bootes & 38$^{+10}_{-9}$ & 289$^{\pm30}$ & 8.9$^{+4.1}_{-3.1}\times10^{-5}$ & 85$^{+20}_{-10}$ & 122$^{\pm15}$ & 5.2$^{+2.7}_{-1.8}\times10^{-4}$ \\
Ursa Major II & 38$^{+3}_{-3}$ & 298$^{\pm19}$ & 1.3$^{+0.7}_{-0.6}\times10^{-4}$ & 99$^{+61}_{-31}$ & 116$^{\pm34}$ & 8.6$^{+12.2}_{-4.6}\times10^{-4}$ \\
Bootes II & 39$^{+1}_{-2}$ & 396$^{\pm70}$ & 2.8$^{+5.3}_{-2.3}\times10^{-4}$ & 537$^{+600}_{-406}$ & 118$^{\pm92}$ & 3.0$^{+14.4}_{-2.6}\times10^{-3}$ \\
Coma Berenices & 42$^{+2}_{-2}$ & 284$^{\pm23}$ & 9.6$^{+3.9}_{-3.2}\times10^{-5}$ & 96$^{+59}_{-28}$ & 125$^{\pm38}$ & 5.0$^{+6.5}_{-2.4}\times10^{-4}$ \\
Sculptor & 60$^{+8}_{-7}$ & 255$^{\pm19}$ & 4.8$^{+1.4}_{-1.3}\times10^{-4}$ & 113$^{+23}_{-12}$ & 134$^{\pm15}$ & 1.8$^{+0.7}_{-0.5}\times10^{-3}$ \\
Sextans & 76$^{+8}_{-8}$ & 265$^{\pm14}$ & 3.2$^{+1.2}_{-1.0}\times10^{-4}$ & 191$^{+126}_{-53}$ & 106$^{\pm25}$ & 2.0$^{+2.0}_{-0.9}\times10^{-3}$ \\
Horologium 1 & 79$^{+12}_{-22}$ & 245$^{\pm37}$ & 1.4$^{+2.1}_{-1.1}\times10^{-4}$ & 129$^{+246}_{-46}$ & 138$^{\pm36}$ & 5.1$^{+10.6}_{-4.1}\times10^{-4}$ \\
Fornax & 79$^{+39}_{-27}$ & 242$^{\pm48}$ & 8.6$^{+4.2}_{-3.1}\times10^{-4}$ & 152$^{+25}_{-6}$ & 120$^{\pm19}$ & 3.5$^{+1.5}_{-1.0}\times10^{-3}$ \\
Carina & 81$^{+24}_{-23}$ & 214$^{\pm37}$ & 3.2$^{+2.2}_{-1.2}\times10^{-4}$ & 106$^{+15}_{-4}$ & 157$^{\pm19}$ & 6.5$^{+3.6}_{-2.4}\times10^{-4}$ \\
Leo I & 101$^{+85}_{-61}$ & 320$^{\pm67}$ & 3.0$^{+0.7}_{-1.1}\times10^{-4}$ & 525$^{+204}_{-27}$ & 138$^{\pm55}$ & 1.6$^{+1.6}_{-0.8}\times10^{-3}$ \\
Ursa Major & 102$^{+6}_{-7}$ & 272$^{\pm49}$ & 2.4$^{+1.4}_{-0.9}\times10^{-4}$ & 408$^{+382}_{-256}$ & 121$^{\pm47}$ & 1.3$^{+1.6}_{-0.7}\times10^{-3}$ \\
Aquarius II & 103$^{+4}_{-7}$ & 433$^{\pm169}$ & 6.2$^{+15.1}_{-5.1}\times10^{-5}$ & 1319$^{+915}_{-1151}$ & 299$^{\pm203}$ & 1.4$^{+9.1}_{-1.2}\times10^{-4}$ \\
Grus I & 116$^{+12}_{-14}$ & 340$^{\pm81}$ & 3.1$^{+6.7}_{-2.5}\times10^{-5}$ & 891$^{+506}_{-526}$ & 161$^{\pm116}$ & 1.4$^{+13.6}_{-1.2}\times10^{-4}$ \\
Hydra II & 137$^{+12}_{-32}$ & 456$^{\pm241}$ & 4.6$^{+15.0}_{-4.2}\times10^{-6}$ & 1488$^{+1402}_{-1053}$ & 353$^{\pm294}$ & 9.3$^{+112.6}_{-8.6}\times10^{-6}$ \\
Canes Venatici I & 138$^{+46}_{-77}$ & 249$^{\pm64}$ & 3.4$^{+0.9}_{-1.3}\times10^{-4}$ & 323$^{+220}_{-64}$ & 106$^{\pm38}$ & 1.9$^{+2.4}_{-0.7}\times10^{-3}$ \\
Leo IV & 154$^{+5}_{-7}$ & 504$^{\pm259}$ & 1.5$^{+4.5}_{-1.2}\times10^{-5}$ & 1799$^{+1349}_{-1269}$ & 424$^{\pm291}$ & 2.3$^{+15.3}_{-1.9}\times10^{-5}$ \\
Canes Venatici II & 159$^{+5}_{-49}$ & 316$^{\pm108}$ & 7.6$^{+8.8}_{-4.2}\times10^{-5}$ & 786$^{+718}_{-545}$ & 142$^{\pm132}$ & 3.5$^{+10.7}_{-2.8}\times10^{-4}$ \\
Leo II & 160$^{+67}_{-115}$ & 226$^{\pm92}$ & 4.0$^{+2.4}_{-2.3}\times10^{-4}$ & 239$^{+258}_{-18}$ & 117$^{\pm42}$ & 1.5$^{+2.8}_{-0.5}\times10^{-3}$ \\
Leo V & 171$^{+6}_{-6}$ & 543$^{\pm257}$ & 1.4$^{+5.2}_{-1.2}\times10^{-5}$ & 1985$^{+1258}_{-1213}$ & 473$^{\pm289}$ & 2.0$^{+13.1}_{-1.8}\times10^{-5}$ \\
Pisces II & 182$^{+15}_{-15}$ & 687$^{\pm375}$ & 2.3$^{+8.4}_{-1.9}\times10^{-5}$ & 2662$^{+1751}_{-1419}$ & 636$^{\pm404}$ & 2.9$^{+16.8}_{-2.4}\times10^{-5}$ \\
Eridanus II & 365$^{+16}_{-17}$ & 955$^{\pm471}$ & 1.9$^{+5.0}_{-1.1}\times10^{-5}$ & 3865$^{+2033}_{-1814}$ & 931$^{\pm482}$ & 2.0$^{+6.0}_{-1.2}\times10^{-5}$ \\
Phoenix$^3$ & 417$^{+20}_{-18}$ & 304$^{\pm152}$ & 3.3$^{+6.5}_{-1.9}\times10^{-4}$ & 1306$^{+697}_{-440}$ & 255$^{\pm165}$ & 4.5$^{+17.8}_{-2.9}\times10^{-4}$ \\
\end{longtable*}
\end{ThreePartTable}
\end{center}

\bibliographystyle{apj}
\bibliography{ref}

\end{document}